RESEARCH

# Rewiring Human Brain Networks via Lightweight Dynamic Connectivity Framework: An EEG-Based Stress Validation

Sayantan Acharya[1] · Abbas Khosravi[1] · Douglas Creighton[1] · Roohallah Alizadehsani[1,*] · U. Rajendra Acharya[2]

**Abstract**
In recent years, Electroencephalographic (EEG) analysis has gained prominence in stress research when combined with AI and Machine Learning (ML) models for validation. In this study, a lightweight dynamic brain connectivity framework based on Time-Varying Directed Transfer Function (TV-DTF) is proposed, where TV-DTF features were validated through ML-based stress classification. TV-DTF estimates the directional information flow between brain regions across distinct EEG frequency bands, thereby capturing temporal and causal influences that are often overlooked by static functional connectivity measures. EEG recordings from the 32-channel SAM 40 dataset were employed, focusing on mental arithmetic task trials. The dynamic EEG-based TV-DTF features were validated through ML classifiers such as Support Vector Machine (SVM), Random Forest (RF), Gradient Boosting (GB), Adaptive Boosting (AdaBoost), and Extreme Gradient Boosting (XGBoost). Experimental results show that alpha-TV-DTF provided the strongest discriminative power, with SVM achieving 89.73% accuracy in 3-class classification and with XGBoost achieving 93.69% accuracy in 2-class classification. Relative to absolute power and phase-locking–based functional connectivity features, alpha-TV-DTF and beta-TV-DTF achieved higher performance across the ML models, highlighting the advantages of dynamic over static measures. Feature importance analysis further highlighted dominant long-range frontal–parietal and frontal–occipital informational influences, emphasizing the regulatory role of frontal regions under stress. These findings validate the lightweight TV-DTF as a robust framework, revealing spatiotemporal brain dynamics and directional influences across different stress levels.

## Introduction

Cognitive stress is defined by an escalating psycho-physiological (psy-phy) response that takes place within the brain when individuals are exposed to compulsion or apprehension while encountering a mentally challenging situation [1]. These situations could be physical, which is due to illness or acute wounds, socially caused by severe financial difficulties or poor living, emotional, which is attributable to job loss or a person's demise, or traumatic, which occurs through disturbing past experiences. Chronic stress is severely harmful, which happens from long-established stressors that imply cardiovascular diseases [2], Post Traumatic Stress Disorder (PTSD) [3] and substance abuse [4]. In general, mental stress has become a progressive aspect that can cause several psychological and physical health complications, which makes it imperative to understand and classify it.

Stress perception is highly individual and idiosyncratic. This contributing factor itself poses a challenge for classifying stress perception levels, as accurate measurement significantly depends on the assessment methodology. Researchers have evaluated mental stress using two primary methods: psychological questionnaires and physiological markers. Among existing psychological questionnaires, the Perceived Stress Scale (PSS) is a competent choice for evaluating stress [5]. However, psychological questionnaires alone cannot provide a comprehensive evaluation of stress, as stress is a dynamic "fight or flight" response originating in the brain and involving multiple physiological functions. Hence, researchers utilize subjective questionnaires with

✉ Name: Roohallah Alizadehsani
Email: r.alizadehsani@deakin.edu.au

1 Institute for Intelligent Systems Research and Innovation (IISRI), Deakin University, Waurn Ponds, VIC, Australia.

2 University of Southern Queensland, Queensland, Australia.

physiological markers such as EEG [6], heart rate variability [7], skin conductance [8], salivary cortisol [9] and blood pressure [10].

Nevertheless, researchers have evidenced that mental stress is centrally administered by the Sympathetic Nervous System (SNS), which is associated with brain regions such as the Hypothalamus, Pituitary gland, and Adrenal cortex collectively identified as the HPA axis [11]. EEG is considered in this study to gain insight into stress-related dynamic changes in cognitive processes, as it's the most effective brain signal modality for stress measurement due to its capability for supporting brain frequency-specific analysis, providing event-related potential patterns, and capturing moment-by-moment functional and connectivity variations of multiple brain regions [6] [12] [13].

EEG is a complex signal that can be non-invasively recorded utilizing exterior electrodes placed on the scalp. Setting up an EEG is economical and user-friendly. The EEG spectrum used in this study is allocated into 5 spectral bands as provided in Table 1. Each of these EEG bands signifies a state of mind of an individual [14]. Heightened beta EEG spectrum power correlates with increased alertness and arousal, while elevated alpha activity is linked to relaxation, and theta waves are characteristic of the sleep state [15].

**Table 1** EEG Spectrum used in this study.

| Delta | Theta | Alpha | Beta | Gamma |
|---|---|---|---|---|
| 0.5-4 Hz | 4-8 Hz | 8-13 Hz | 13-30 Hz | 30-45 Hz |

Beforehand, most of the stress studies utilized Power Spectral Density (PSD) based metrics of frontal EEG channels [12] [16]. So far, the highest accuracy achieved in classifying stress is 99.94%, 99.93%, and 99.75% utilizing classifiers such as Linear Discriminant Analysis (LDA), k-Nearest Neighbor (kNN) and cubic SVM respectively, by employing PSD-based features such as Median Frequency, Spectral Moments, Root Mean Square (RMS) and Modified Frequency Mean features over relax-stress condition [17]. However, the stress-building process is closely coupled with the brain regions, as it's the brain that perceives a threat or a challenge and initiates the stress response [12] [16] [18]. Hence, utilizing brain connectivity analysis may provide operative insights into the functional and effective variations that may bestow an accurate model of brain activity concerning how its diverse regions communicate and influence each other during the stress-building process [19].

Previously, in brain connectivity, researchers have employed Magnitude Square Coherence (MSC), Coherence, Phase–slope Index (PSI), Canonical Correlation Analysis (CCA), and Mutual Information (MI) to classify stress. Extracting MSC features has provided superior accuracy by applying SVM compared to PSI, CCA, and PSD [12].

Nonetheless, the primary disadvantage of MSC and coherence analysis is its elevated susceptibility to power variations and phase coupling [20] [21]. This elevated susceptibility to phase coupling and power changes can result in detecting connectivity even when there might not be a true functional relationship between the signals. This can lead to identifying spurious connections, and true connectivity patterns may be obscured [22]. On the contrary, CCA is beneficial when using cross-covariance matrices to evaluate the influence of psychological stress [23]. However, CCA is limited to linear relationships [23]. It may not capture complex nonlinear associations or interactions that exist in the signal. There is a study where researchers have utilized PSI, general PDC (gPDC) features in classifying stress, where PSI has achieved low accuracy when compared to PDC and Directed Transfer Function (DTF) [18]. On the other hand, PSI is primarily used to measure phase synchronization, where it may fail to appropriately identify any directional causal influence of information flow among brain regions [24]. Moreover, previously used DTF features were not time-varying, meaning they do not provide information about the causal influence among the EEG channels at specific temporal windows, potentially missing out significant data about the brain dynamics. Concurrently, researchers have used Phase Locking Value (PLV) to estimate functional connectivity features; however, it provides static information about the phase synchronization between two brain regions [25]. Similar to PLV, another article has been found where researchers have implemented Amplitude Envelope Correlation (AEC). This functional connectivity measure does not provide the variability of time windows and the directional influence of brain regions [26]. In summary, the contributions of this paper are listed as follows.:

- A lightweight framework is proposed for computing the Time-Varying Directed Transfer Function (TV-DTF).
- A Machine Learning (ML)-based validation of TV-DTF features was performed for mental stress quantification through 2-class and 3-class classification using SVM, RF, GB, AdaBoost, and XGBoost, following the comparison with PSD and PLV-based 3-class classification scenarios.
- Evaluation of top-ranked TV-DTF features to examine the dynamic connectivity patterns and their impact on ML-based stress classification.

This paper is structured as Materials presents the EEG-based materials and stress task, the computation method in Lightweight Time-Varying Directed Transfer Function Framework followed by Machine Learning based Validation results, including the evaluation and impact of TV-DTF features, and Discussion. Lastly, the onclusion is provided at the end of the paper.

## Materials

The EEG recordings from the SAM 40 dataset [27] were used in this study to evaluate time-varying brain connectivity under stress. The dataset provides multichannel signals labelled into three stress levels, such as Relaxed, Low Stress, and High Stress, derived from task performance and participant responses. Among the stress task-oriented experimental protocols, the Mental Arithmetic Task (MAT) was selected as it has been extensively validated for inducing cognitive stress, particularly in regions governing attention, executive control, and working memory [28].

### EEG Dataset

The SAM 40 dataset contains EEG recordings of 40 participants (26 males, 14 females, mean age of 21.5), sample frequency of 128 Hz [27]. It was recorded while the participants were performing 3 stress tasks, such as distinguishing Symmetric Mirror Images (SMI), solving the Stroop Colour-Word Task (SCWT), and completing Mental Arithmetic. The EEG signals were captured utilizing a 32-channel Emotiv Epoc Flex gel kit [27]. Nevertheless, the researchers have pre-processed the EEG data to eliminate the baseline drifts by subtracting the average trend attained through the Savitzky-Golay filtering method [27]. Physiological noises were also eliminated by employing wavelet thresholding [27].

### Stress Task

The MAT-based EEG data from the SAM 40 dataset have been considered for time-varying analysis [27]. The MAT has been widely recognized in psy-phy research for reliably inducing stress-related changes in brain activity, particularly in regions associated with executive function, attention, and cognitive control [28]. In the dataset, participants are instructed to mentally solve arithmetic problems and indicate their response with a thumbs-up or thumbs-down gesture based on whether the solution shown on the screen is correct or not [27].

As provided in Fig. 1 a total of six different arithmetic problems involving various operators are presented to each participant in every single trial during the experiment, where each trial provides 25 seconds of EEG data. The choice of MAT is also driven by the consistency of EEG alterations observed across participants, whereas the two other tasks, SCWT and recognition of SMI, did not induce noticeable alterations in the cerebral activities as captured by EEG signals.

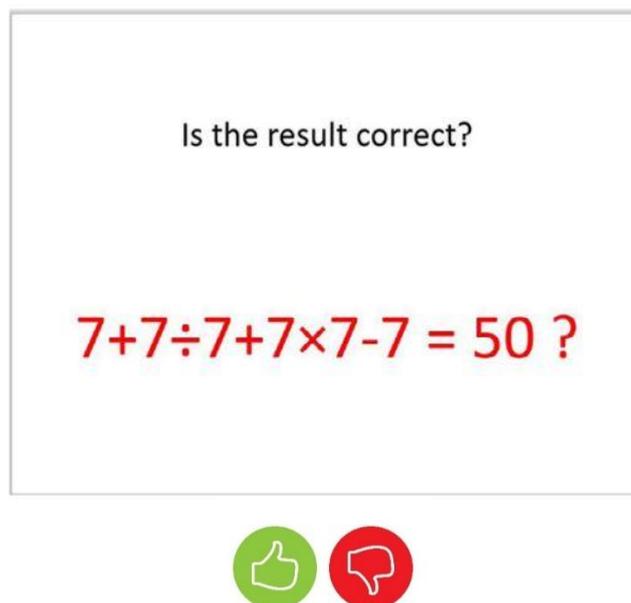

**Fig. 1** Structure of the stress task [27].

### Data Labelling

A self-reported stress rating approach, based on a 10-point Likert scale, was employed to label EEG trials as High Stress or Low Stress conditions according to participants' feedback during the MAT. For each participant, three MAT trials were recorded, and the highest-rated trial (e.g., 8/10) was labelled as High Stress, while the lowest-rated trial (e.g., 3/10) was labelled as Low Stress. The database also contains 3 distinct Relaxed data taken before each trial of MAT for all 40 subjects [27]. In addition, the dataset included three distinct Relaxed trials recorded before each MAT session. To ensure reliable labelling, data from five participants were excluded due to inconsistent or non-discriminative ratings, leaving EEG data from 35 participants for investigation.

## Lightweight Time-Varying Directed Transfer Function Framework

Directed Transfer Function (DTF) is a well-recognised Granger Causality-based technique employed in EEG analysis to examine the effective connectivity among cerebral cortices [29]. In this study, it is applied to estimate the directional influences and information flow among EEG electrodes associated with the distinctive brain regions. The primary advantage of utilising DTF is that it allows researchers to examine the frequency-specific connectivity and directed information flow between different electrodes [29]. DTF is derived from the frequency-domain transfer function of the MVAR model and is calculated as the normalised square magnitude of the transfer function between brain regions [29].

By estimating the frequency-domain transfer function of the MVAR model, DTF quantifies the magnitude (strength) of the overall directed influence from one electrode to another at different frequencies. DTF could be calculated by estimating Multivariate Autoregressive (MVAR) models [30]. Numerous frequency-domain connectivity estimators have been applied via MVAR modelling of the EEG time series data. A time-varying MVAR model with $N$ EEG signal channels of model order $p$ is specified with [30],

$$X(t) = \sum_{r=1}^{p} A_r(t) X(t-r) + E(t) \quad (1)$$

Here, $X(t) = [x_1(t), x_2(t), \ldots, x_N(t)]^*$, $x_j$ denotes the $j^{th}$ signal source at time point $t$ where $t = 1, 2, \ldots, L$ where $L$ is the total no. of time points. The asterisk * denotes the transpose of a matrix. $A_r(t)$ signifies the $N \times N$ model coefficient matrix at lag $r = 1, \ldots, p$. $E(t) = [e_1(t), e_2(t), \ldots, e_N(t)]^*$ indicates the approximated error, which is a vector of 0 mean input of white noise with a covariance matrix $\Sigma_E$.

Hence, the time-varying MVAR model could be represented as,

$$X(t) = \sum_{r=1}^{p} A(r, t) X(t-r) + E(t) \quad (2)$$

Where $A(r, t)$ is the MVAR model coefficient matrix at time point $t$ with lag $r$.

To determine the approximation error, Eq. (2) can be rearranged as follows:

$$E(t) = \sum_{r=1}^{p} \hat{A}(r) X(t-r) \quad (3)$$

Where $\hat{A}(r)$ is,

$$\hat{A}(r) = \begin{cases} 1 - A(r), & \text{for } r = 0 \\ -A(r), & \text{for } r > 0 \end{cases} \quad (4)$$

The frequency-domain characterization of Eq. (1) could be provided by altering the time-domain first into the $z$ domain, after that changing from the $z$ domain to the frequency-domain, such as,

$$A(z, t) = I - \sum_{r=1}^{p} A_r(t) z^{-r},$$

$$H(z, t) = A^{-1}(z, t), \quad z = e^{-j2\pi f/fs}$$

$$H(z, t) = \frac{X(z,t)}{E(z,t)}$$

$$H(f, t) = \frac{X(f,t)}{E(f,t)} = A^{-1}(f, t) \quad (5)$$

Here, $H(f, t)$ signifies the system transfer function, $E(f, t)$ is the system input, and $X(f, t)$ is the system output.

Then, the frequency-domain representation is,

$$A(f, t) = I - \sum_{r=1}^{p} A_r(t) e^{-j2\pi fr/fs} \quad (6)$$

Thereby, from Eq. (5), the time-varying DTF could be calculated by employing the following equation [161],

$$DTF_{i,j}(f, t) = \frac{|H_{i,j}(f, t)|}{\sqrt{\sum_{m=1}^{M} |H_{mi}(f, t)|^2}}$$

This could be rewritten as,

$$DTF_{i,j}(f, t) = \frac{|H_{i,j}(f, t)|}{\sqrt{h_i^*(f, t) h_i(f, t)}} \quad (7)$$

Where $H_{i,j}(f, t)$ is the $(i, j)$ entry of the transfer matrix $H(f, t)$, and $h_i(f, t)$ denotes the $i^{th}$ column of $H(f, t)$, so $h_i^*$ is its conjugate transpose. $DTF_{i,j}(f, t)$ quantifies the magnitude and the direction of the information flow commencing in channel $j \rightarrow i$ at frequency $f$ and at time-window $t$.

The TV-DTF analysis was carried out by developing a novel connectivity framework in MATLAB (version R2022b). The analysis was computed in 4 steps to estimate the TV-DTF features from EEG channels, all frequency bands, as provided in Table 1. The first step in the TV-DTF estimation is to load the EEG data by selecting the EEG channels and key parameters, such as window size, step size, and sampling frequency. In this research, all 32 EEG channels were selected from the MAT stimulus of the SAM 40 dataset with a window size of 5 seconds, and a step size of 2 seconds, while the sampling frequency of 128Hz was provided.

After loading the EEG data by defining the no. of channels and key parameters, the EEG data is segmented into smaller chunks based on the window size and step size, also known as the Short Time Windowing (STW) technique [31] [32]. This segmentation technique is mandatory to examine the temporal dynamics of the brain, which allows us to track how the brain connectivity patterns evolve over time [31]. Short 5-second windows were selected to investigate the temporal dynamics in the connectivity patterns. Previously, one study utilized 5-second windows while implementing the STW technique [32]. The overlapping step size of 2 seconds was selected, which is typically half of the time windows.

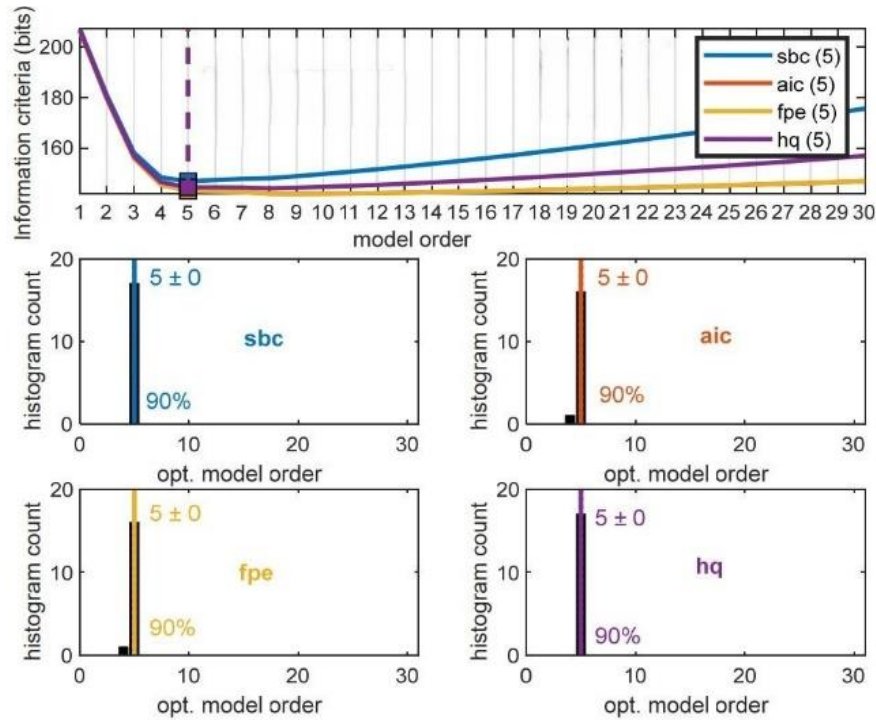

**Fig. 2** Model order selection using AIC, SBC, FPE and HQ

After segmentation, the EEG segments were extracted by computing the start and end indices of the corresponding segments. The subsequent critical step is calculating the model order p (see Eq. (1)). The model order represents the number of previous time points of each EEG channel, used to predict its future values [33]. Selecting an appropriate model order is crucial to maintaining balance, which is required to accurately capture the underlying brain dynamics while preventing overfitting or underfitting [34]. From Fig. 2, the model order 5 was selected based on the optimum model order selection result by calculating Akaike's Final Prediction Error criterion (FPE), Hannan–Quinn criterion (HQ), Akaike's Information Criterion (AIC), and Schwarz–Bayes Criterion (SBC) by using ARfit toolbox along with the aid of Source Information Flow Toolbox (SIFT) extension in EEGLAB [33] [35] [36]. The formulations of AIC, SBC, FPE and HQ can be found in [36].

After estimating the optimized model order p, the next step involves fitting or adapting the MVAR model to the EEG channels (see Eq. (2)). This step captures the dynamic relationships and causal influences between different EEG channels. In this study, the Yule-Walker approach is employed to calculate the MVAR model coefficients [37]. The model fitting typically refers to estimating the coefficients that describe how past values of the time series influence the current EEG signal.

Along with the defined sampling frequency, the total frequency range is divided into several frequency bins for the connectivity analysis. These bins represent the intervals of the total frequency range used to analyse brain connectivity at specific frequency bands. Consequently, frequency indices are determined, which are used to extract the corresponding frequency bins for each EEG frequency band.

After that, the frequency response is computed for each pair of EEG channels based on the frequency bins. The frequency response computation helps to quantify how activity in each channel influences other EEG channels across different frequency bands, a critical step in the analysis. The TV-DTF feature maps per time window are provided in Fig. 5.

**Algorithm 1** Lightweight Time-Varying Directed Transfer Function (DTF) Computation

**Require:** EEG signal $X \in \mathbb{R}^{T \times N}$, sampling rate $F_s$, model order $P$, window size $L$, overlap $O$, number of frequency bins $F$
**Ensure:** Time-varying DTF tensor $\mathcal{D} \in \mathbb{R}^{N \times N \times W}$, where $W$ is the number of windows

1: Segment $X$ into $W$ overlapping windows of length $L$ and overlap $O$.
2: **for** $t = 1$ to $W$ **do**
3:    $X^{(t)} \leftarrow$ EEG segment of shape $L \times N$.
4:    Estimate MVAR coefficients $\{A_p(t)\}_{p=1}^P$ from $X^{(t)}$.
5:    **for** $m = 0$ to $F - 1$ **do**
6:      $f_m \leftarrow \frac{m F_s}{F}$
7:      $A(f_m, t) \leftarrow I - \sum_{p=1}^{P} A_p(t) e^{-j 2\pi f_m p / F_s}$.
8:      $H(f_m, t) \leftarrow A(f_m, t)^{-1}$
9:      **for** $i = 1$ to $N$ **do**
10:         $D_i(f_m, t) \leftarrow \sum_{k=1}^{N} |H_{ik}(f_m, t)|^2$
11:         **for** $j = 1$ to $N$ **do**
12:            **if** $D_i(f_m, t) > 0$ **then**
13:               $DTF_{ij}(f_m, t) \leftarrow \frac{|H_{ij}(f_m, t)|^2}{D_i(f_m, t)}$.
14:            **else**
15:               $DTF_{ij}(f_m, t) \leftarrow 0$.
16:            **end if**
17:         **end for**
18:      **end for**
19:    **end for**
20:    $\mathcal{D}(:,:,t) \leftarrow \frac{1}{|\mathcal{B}|} \sum_{m \in \mathcal{B}} DTF(:,:,f_m, t)$ *(band-averaged over target band $\mathcal{B}$).*
21: **end for**
22: **Return** $\mathcal{D}$ *(row $i$ sums inflows into target $i$; $DTF_{ij}(f_m, t)$ quantifies $j \to i$).*

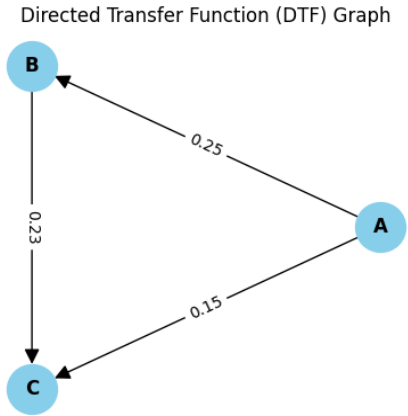

**Fig. 3** Three-node example of the lightweight TV-DTF framework.

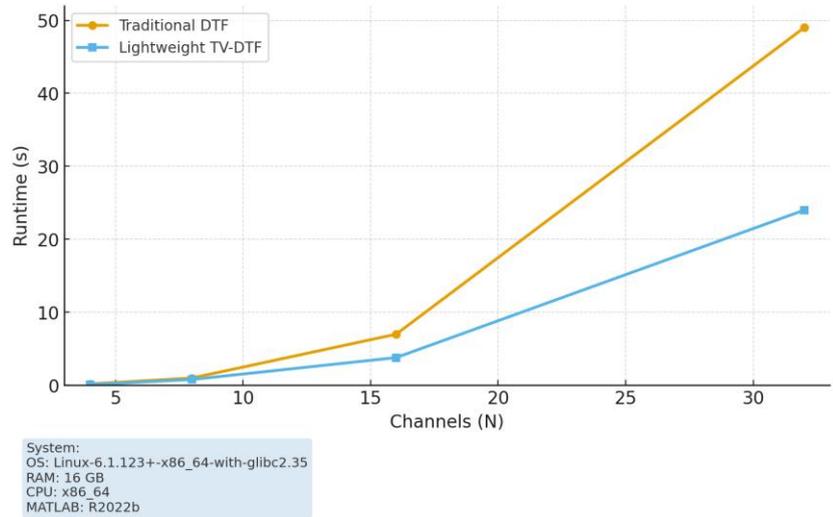

**Fig. 4** Runtime analysis of the lightweight TV-DTF framework.

To demonstrate the efficacy of the lightweight TV-DTF method, a simple system with three nodes, A, B, and C, was used. The arrowheads show the direction of information flow between the nodes, and the numbers on the arrows represent the strength of the influence. In Fig. 3, A influences B with a strength of 0.25, B influences C with a strength of 0.23, and A also influences C with a strength of 0.15. This shows that the lightweight TV-DTF method can successfully detect the expected connections before applying it to larger and complex EEG datasets.

In Fig. 4, the runtime is averaged over frequency bands and temporal windows across 35 participants. The traditional DTF was computed using the SIFT toolbox extension in EEGLAB, which implements multivariate autoregressive modelling for directed effective connectivity analysis [36]. When comparing runtime against the number of channels, it can be observed that the lightweight TV-DTF method scales efficiently with increasing EEG channels. Both the traditional DTF and the proposed lightweight TV-DTF were executed on the same machine with Linux Kernel 6.1.123+, 16 GB RAM, using MATLAB (version R2022b) to ensure a fair runtime comparison.

## Machine Learning based Validation

For the Machine Learning (ML)-based validation phase, the TV-DTF features of 35 participants in all EEG bands, as mentioned in the Table 1, were analysed using subject-wise 10-fold cross-validation, where the data is split based on subjects rather than individual segments. It is crucial to avoid data leakage by ensuring that data from the same participant does not appear in both the training and testing sets. This approach not only prevents data leakage but also ensures that no single subject's data will appear in both the training and test sets simultaneously [38]. For classification purposes, the feature matrices were flattened before being integrated into classifiers, while preserving the magnitude and significance of individual data points.

In this experiment, five ML techniques, such as Support Vector Machine (SVM) [17], Random Forest (RF) [39], Extreme Gradient Boosting (XGBoost) [40], Gradient Boosting (GB) [40], and Adaptive Boosting (AdaBoost) [41], have been used for stress classification. These classifiers were selected to leverage their inherent capabilities of diverse learning mechanisms, robustness in handling high-dimensional data, and suitability for limited datasets. Previously, SVM and Decision Tree (DT) have been used for handling EEG-based PDC features, which are similar to DTF, in decoding EEG-based motor imagery [42]. RF, GB, AdaBoost, and XGBoost algorithms simultaneously excel in capturing non-linear interactions among EEG channels, making them particularly suitable for analyzing EEG-based DTF features [39] [40] [41]. Precision, recall, F1 score, and accuracy were utilized for assessing the effectiveness of the ML models. Precision estimates the ratio of true positives out of all positive instances, reflecting the model's effectiveness in identifying positive cases [39]. Recall indicates the ratio of true positives among all actual positive predictions. The F1 score is the harmonic average of precision and recall, which offers a balanced single metric [41]. Accuracy represents the ratio of true positive and true negative instances out of all predictions. The formulations of performance metrics from Eq. (8) to Eq. (11) are provided below [39] [40] [41].

$$\text{Precision} = \frac{T_P}{T_P + F_P} \qquad (8)$$

$$\text{Recall} = \frac{T_P}{T_P + F_N} \qquad (9)$$

$$\text{F1 score} = 2 \times \frac{Precision \times Recall}{Precision + Recall} \qquad (10)$$

$$\text{Accuracy} = \frac{T_P + T_N}{T_P + T_N + F_P + F_N} \qquad (11)$$

Where $T_P$ presents the true positives, $T_N$ presents the true negatives, $F_P$ presents the false positives and $F_N$ presents the false negatives.

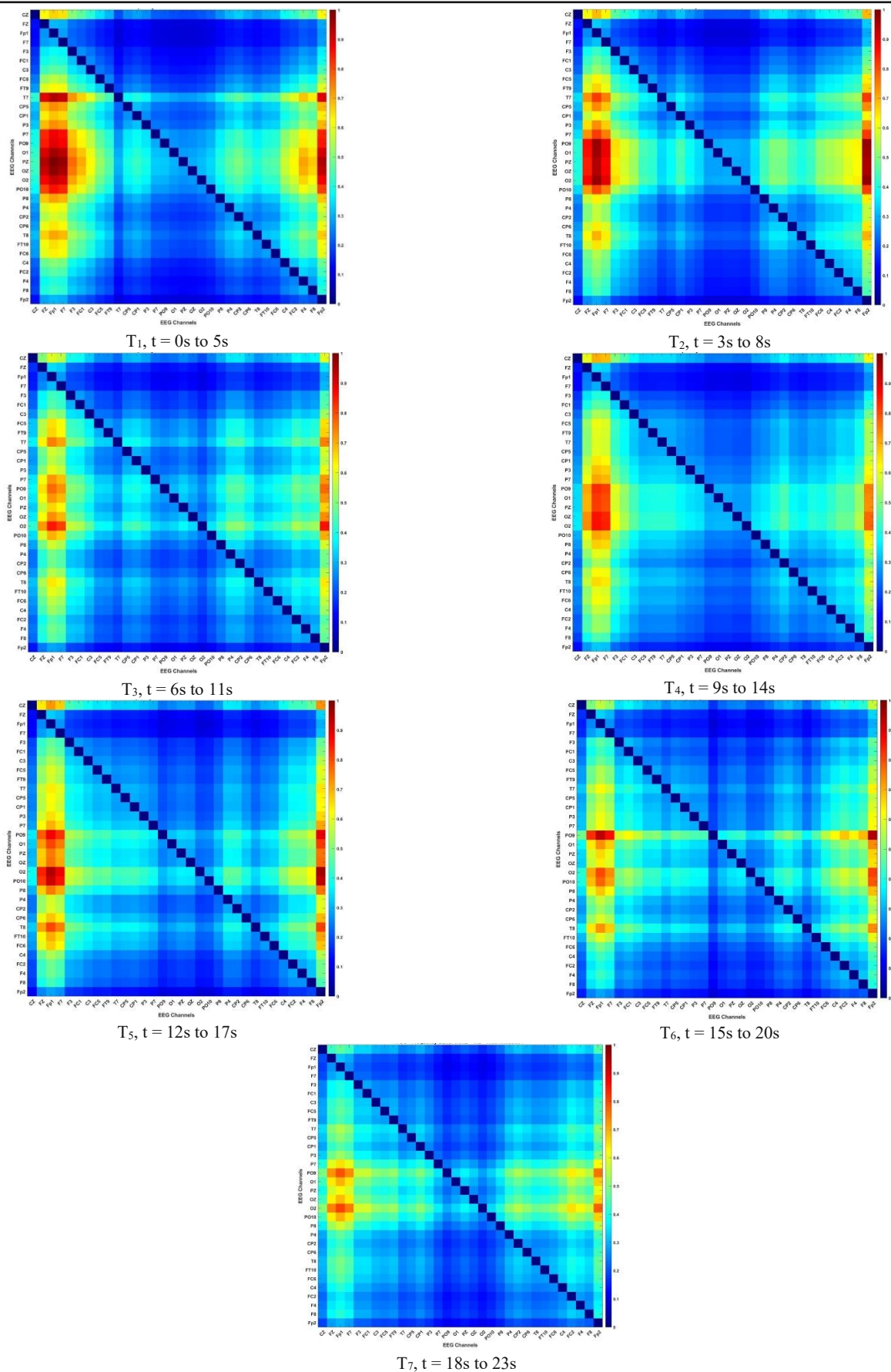

**Fig. 5** Variations of dynamic TV-DTF across seven temporal windows in the alpha band.

Table 2 3-class stress classification results of TV-DTF features in all EEG bands.

| ML Models | Feature Types | Precision (%) (m ± std) | Recall (%) (m ± std) | F1 Score (%) (m ± std) | Accuracy (%) (m ± std) |
|---|---|---|---|---|---|
| SVM | Delta-TV-DTF | 68.43 ± 0.34 | 65.76 ± 1.88 | 67.07 ± 3.05 | 68.26 ± 2.82 |
|  | Theta-TV-DTF | 64.64 ± 2.50 | 53.41 ± 4.25 | 58.49 ± 1.71 | 59.56 ± 2.16 |
|  | Alpha-TV-DTF | 85.12 ± 1.89 | 90.35 ± 2.70 | 87.66 ± 1.37 | 89.73 ± 1.66 |
|  | Beta-TV-DTF | 88.31 ± 1.83 | 80.26 ± 2.45 | 84.09 ± 1.74 | 87.95 ± 1.50 |
|  | Gamma-TV-DTF | 56.26 ± 1.10 | 63.55 ± 2.52 | 59.68 ± 4.55 | 62.64 ± 3.42 |
| RF | Delta-TV-DTF | 51.57 ± 2.73 | 57.95 ± 1.34 | 54.57 ± 4.85 | 56.58 ± 3.95 |
|  | Theta-TV-DTF | 64.23 ± 4.09 | 52.48 ± 1.04 | 57.76 ± 1.35 | 58.95 ± 2.25 |
|  | Alpha-TV-DTF | 73.68 ± 3.59 | 75.11 ± 3.74 | 74.39 ± 3.42 | 76.55 ± 2.29 |
|  | Beta-TV-DTF | 80.13 ± 3.11 | 70.25 ± 3.20 | 74.84 ± 2.79 | 78.43 ± 1.87 |
|  | Gamma-TV-DTF | 66.82 ± 3.78 | 50.26 ± 2.05 | 57.37 ± 3.47 | 59.20 ± 2.88 |
| GB | Delta-TV-DTF | 62.29 ± 3.10 | 52.00 ± 2.61 | 56.68 ± 2.79 | 59.50 ± 3.11 |
|  | Theta-TV-DTF | 55.81 ± 3.89 | 60.20 ± 0.10 | 57.92 ± 0.75 | 59.61 ± 4.64 |
|  | Alpha-TV-DTF | 65.33 ± 3.52 | 75.27 ± 3.90 | 69.95 ± 3.39 | 71.11 ± 2.56 |
|  | Beta-TV-DTF | 67.37 ± 2.26 | 70.00 ± 2.45 | 68.66 ± 1.90 | 71.67 ± 2.55 |
|  | Gamma-TV-DTF | 55.64 ± 0.16 | 54.52 ± 1.15 | 55.07 ± 3.51 | 55.97 ± 1.19 |
| AdaBoost | Delta-TV-DTF | 49.70 ± 2.69 | 54.17 ± 1.13 | 51.84 ± 1.18 | 52.58 ± 3.48 |
|  | Theta-TV-DTF | 56.17 ± 4.87 | 55.21 ± 0.74 | 55.69 ± 3.07 | 57.43 ± 1.12 |
|  | Alpha-TV-DTF | 53.26 ± 3.94 | 65.32 ± 4.36 | 58.68 ± 4.02 | 60.49 ± 2.79 |
|  | Beta-TV-DTF | 70.32 ± 3.59 | 68.77 ± 3.90 | 69.54 ± 3.48 | 70.52 ± 2.25 |
|  | Gamma-TV-DTF | 52.72 ± 0.34 | 59.44 ± 1.28 | 55.88 ± 2.89 | 58.52 ± 4.66 |
| XGBoost | Delta-TV-DTF | 66.62 ± 4.16 | 59.03 ± 2.28 | 62.59 ± 2.68 | 65.27 ± 0.61 |
|  | Theta-TV-DTF | 50.76 ± 2.12 | 63.92 ± 1.48 | 56.58 ± 3.43 | 58.41 ± 1.84 |
|  | Alpha-TV-DTF | 78.37 ± 3.02 | 75.12 ± 3.35 | 76.71 ± 2.66 | 79.38 ± 1.94 |
|  | Beta-TV-DTF | 85.15 ± 2.08 | 80.67 ± 2.45 | 82.85 ± 1.81 | 86.40 ± 1.75 |
|  | Gamma-TV-DTF | 62.79 ± 3.35 | 60.21 ± 0.53 | 61.47 ± 0.10 | 63.22 ± 3.64 |

The classification outcomes, as given in Table 2, delta-TV-DTF and theta-TV-DTF features show modest performance across the ML models in distinguishing stress levels. For Delta, SVM performs best with a precision of 68%, a recall of 66%, an F1 score of 67%, and an accuracy of 68%, making it the strongest model for this band. XGBoost follows with a precision of 67%, a recall of 59%, an F1 score of 64%, and an accuracy of 61%. In comparison, RF and GB show relatively weaker results, with accuracies around 62% to 70%, while AdaBoost yields the lowest precision of 50% but reaches 63% accuracy. For theta-TV-DTF, performances are slightly more varied: SVM achieves 65% precision and 67% accuracy, while RF and GB record moderate accuracies around 56% to 62%. Generally, delta and theta bands provide lower classification power, with SVM consistently outperforming the other models, although their results remain weaker than higher frequency bands.

The alpha-TV-DTF and beta-TV-DTF features demonstrate strong predictive performance across ML models, with Alpha providing the highest classification scores overall. SVM excels in the alpha band, achieving 85% precision, 90% recall, 88% F1 score, and 90% accuracy, establishing it as the most reliable model for this frequency range. RF also performs competitively in alpha, with 74% precision, 75% recall, an F1 score of 72%, and an accuracy of 70%, while GB and AdaBoost remain moderate at 65 % to 70% accuracy. In the beta band, both SVM and XGBoost emerge as top performers. SVM records 88% precision, 80% recall, F1 score of 84%, and an accuracy of 88%, while XGBoost closely follows with 85% precision, 81% recall, F1 score of 79%, and an accuracy of 88%. RF achieves a respectable 80% precision and 73% accuracy, whereas AdaBoost trails slightly with an accuracy of 70%. These results highlight alpha and beta bands as the most discriminative for stress classification, with SVM and XGBoost consistently producing superior outcomes.

The gamma-TV-DTF features produce comparatively lower classification outcomes across all ML models, though some models still yield moderate performance. RF demonstrates the highest gamma precision at 67%, but accuracy remains around 56%. XGBoost also performs moderately with 63% precision, 60% recall, 52% F1 score, and 59% accuracy, while SVM records 56% precision, 64% recall, 60% accuracy, and 60% F1 score. Overall, gamma provides less discriminative information compared to Alpha and Beta, leading to weaker classification performance. When comparing across all bands, alpha-TV-DTF provides the best classification outcomes, with SVM emerging as the most effective model overall, followed closely by XGBoost in beta. RF and GB show balanced but moderate outcomes, while AdaBoost consistently underperforms across most bands. These findings suggest that alpha and beta connectivity patterns are the most robust for identifying stress levels, and SVM in particular demonstrates superior generalisation across frequency bands.

Next, the 3-class TV-DTF-based classification performances were evaluated through a comparative analysis with the 3-class stress scenarios using PSD-based Absolute Power (AP) and Phase Locking Value (PLV), a functional connectivity metric of alpha and beta bands. Alpha and beta-based TV-DTF features demonstrated the highest discriminative ability in stress classification; therefore, these bands were selected for comparison against AP and PLV-based results. PSD is a function that quantifies the power distribution of a time-series signal per unit frequency [8] [9] [11]. In this research, Welch's method is used to calculate the PSD [16]. Welch's method for calculating PSD involves dividing the EEG signal into overlapping segments, each processed with a window function, such as Hamming or Hann, to minimize the spectral leakage [9] [16] [11].

Table 3 Comparison of alpha-AP and alpha-PLV with alpha-TV-DTF in 3-class scenario.

| ML Models | Feature Type | Precision (%) (m ± std) | Recall (%) (m ± std) | F1 Score (%) (m ± std) | Accuracy (%) (m ± std) |
|---|---|---|---|---|---|
| SVM | Alpha-AP | 66.27 ± 3.50 | 60.54 ± 2.90 | 63.66 ± 3.10 | 68.50 ± 3.20 |
|  | Alpha-TV-DTF | 85.12 ± 1.89 | 90.35 ± 2.70 | 87.66 ± 1.37 | 89.73 ± 1.66 |
|  | Alpha-PLV | 76.10 ± 3.18 | 70.85 ± 2.85 | 73.38 ± 2.40 | 75.21 ± 2.32 |
| RF | Alpha-AP | 58.51 ± 1.55 | 58.64 ± 4.32 | 58.57 ± 2.94 | 58.48 ± 2.94 |
|  | Alpha-TV-DTF | 73.68 ± 3.59 | 75.11 ± 3.74 | 74.39 ± 3.42 | 76.55 ± 2.29 |
|  | Alpha-PLV | 73.45 ± 2.25 | 68.92 ± 2.70 | 71.11 ± 2.20 | 72.84 ± 2.15 |
| GB | Alpha-AP | 51.83 ± 4.76 | 51.29 ± 0.82 | 51.56 ± 2.79 | 52.20 ± 2.70 |
|  | Alpha-TV-DTF | 65.33 ± 3.52 | 75.27 ± 3.90 | 69.95 ± 3.39 | 71.11 ± 2.56 |
|  | Alpha-PLV | 70.12 ± 2.60 | 66.45 ± 2.95 | 68.23 ± 0.35 | 69.57 ± 2.40 |
| AdaBoost | Alpha-AP | 53.41 ± 3.65 | 58.47 ± 1.54 | 55.83 ± 2.60 | 54.46 ± 2.60 |
|  | Alpha-TV-DTF | 53.26 ± 3.94 | 65.32 ± 4.36 | 58.68 ± 4.02 | 60.49 ± 2.79 |
|  | Alpha-PLV | 66.80 ± 2.90 | 62.73 ± 3.10 | 64.70 ± 2.85 | 66.42 ± 2.75 |
| XGBoost | Alpha-AP | 56.74 ± 2.17 | 53.38 ± 2.49 | 55.01 ± 2.34 | 56.12 ± 2.34 |
|  | Alpha-TV-DTF | 78.37 ± 3.02 | 75.12 ± 3.35 | 76.71 ± 2.66 | 79.38 ± 1.94 |
|  | Alpha-PLV | 78.55 ± 1.95 | 72.91 ± 2.55 | 75.61 ± 2.10 | 77.83 ± 2.05 |

Table 4 Comparison of beta-AP and beta-PLV with beta-TV-DTF in 3-class scenario.

| ML Models | Feature Types | Precision (%) (m ± std) | Recall (%) (m ± std) | F1 Score (%) (m ± std) | Accuracy (%) (m ± std) |
|---|---|---|---|---|---|
| SVM | Beta-AP | 58.49 ± 0.94 | 50.77 ± 1.88 | 54.36 ± 1.36 | 56.42 ± 1.36 |
|  | Beta-TV-DTF | 88.31 ± 1.83 | 80.26 ± 2.45 | 84.09 ± 1.74 | 87.95 ± 1.50 |
|  | Beta-PLV | 78.20 ± 2.05 | 72.65 ± 2.60 | 75.32 ± 2.20 | 76.98 ± 2.10 |
| RF | Beta-AP | 57.01 ± 3.69 | 54.53 ± 0.92 | 55.74 ± 2.31 | 57.20 ± 2.50 |
|  | Beta-TV-DTF | 80.13 ± 3.11 | 70.25 ± 3.20 | 74.84 ± 2.79 | 78.43 ± 1.87 |
|  | Beta-PLV | 76.40 ± 2.15 | 71.58 ± 2.55 | 73.90 ± 2.10 | 75.42 ± 3.02 |
| GB | Beta-AP | 54.34 ± 2.42 | 57.74 ± 1.69 | 55.99 ± 2.06 | 56.39 ± 2.06 |
|  | Beta-TV-DTF | 67.37 ± 2.26 | 70.00 ± 2.45 | 68.66 ± 1.90 | 71.67 ± 2.55 |
|  | Beta-PLV | 73.12 ± 2.40 | 68.44 ± 2.85 | 70.70 ± 2.25 | 72.18 ± 2.15 |
| AdaBoost | Beta-AP | 50.78 ± 4.62 | 57.45 ± 2.45 | 53.91 ± 3.54 | 56.69 ± 3.54 |
|  | Beta-TV-DTF | 70.32 ± 3.59 | 68.77 ± 3.90 | 69.54 ± 3.48 | 70.52 ± 2.25 |
|  | Beta-PLV | 69.20 ± 2.80 | 65.31 ± 3.05 | 67.20 ± 2.70 | 69.85 ± 2.60 |
| XGBoost | Beta-AP | 58.00 ± 4.71 | 54.51 ± 3.11 | 56.20 ± 3.91 | 54.21 ± 3.91 |
|  | Beta-TV-DTF | 85.15 ± 2.08 | 80.67 ± 2.45 | 82.85 ± 1.81 | 86.40 ± 1.75 |
|  | Beta-PLV | 80.35 ± 1.85 | 74.92 ± 2.40 | 77.52 ± 2.08 | 79.61 ± 1.95 |

From Table 3, it can be observed that alpha-TV-DTF consistently outperforms alpha-AP across all ML models, emphasizing the dynamic temporal connectivity patterns over static power features. For instance, SVM with alpha-TV-DTF achieved 85% precision, 90% recall, 88% F1 score, and 90% accuracy, in contrast to alpha-AP, where SVM peaked with 66% precision, 60% recall, 63% F1 score, and 69% accuracy. Similarly, RF with alpha-TV-DTF obtained 74% F1 and 77% accuracy, which is higher than the 59% F1 and 58% accuracy observed with alpha-AP. Among ensemble models, GB and AdaBoost showed modest outcomes with alpha-AP, with GB achieving 51% F1 and 52% accuracy and AdaBoost reaching 56% F1 and 55% accuracy. However, their performance peaked with alpha-TV-DTF, where GB obtained 70% F1 and 71% accuracy, and AdaBoost improved to 59% F1 and 61% accuracy. XGBoost followed the same trend, advancing from 55% F1 and 56% accuracy with alpha-AP to 77% F1 and 79% accuracy with alpha-TV-DTF. These results confirm that temporal dynamics captured by TV-DTF enhance stress classification beyond what static alpha-band power provides.

Similarly, the classification results using beta-AP and beta-TV-DTF reveal clear improvements when temporal features are employed. From Table 4, beta-TV-DTF outclasses beta-AP in all ML models, highlighting its ability to detect stress-related variations with fewer misclassifications. For example, SVM with beta-TV-DTF achieved 88% precision, 80% recall, 84% F1 score, and 88% accuracy, while beta-AP with SVM provided 58% precision, 51% recall, 54% F1 score, and 56% accuracy. Likewise, RF with beta-TV-DTF reached 75% F1 and 78% accuracy, compared to beta-AP's 56% F1 and 57% accuracy. GB models further reinforce this trend, improving from 56% F1 and 57% accuracy with beta-AP to 69% F1 and 72% accuracy with beta-TV-DTF.

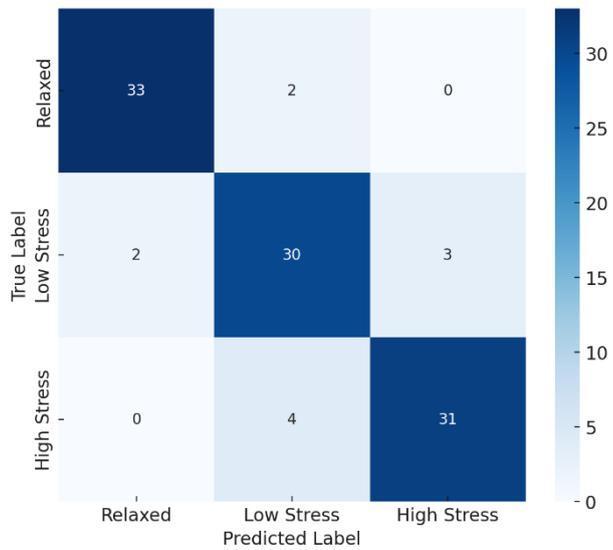

**Fig. 6** Confusion matrix of raw counts for 3-class classification using alpha-TV-DTF with SVM.

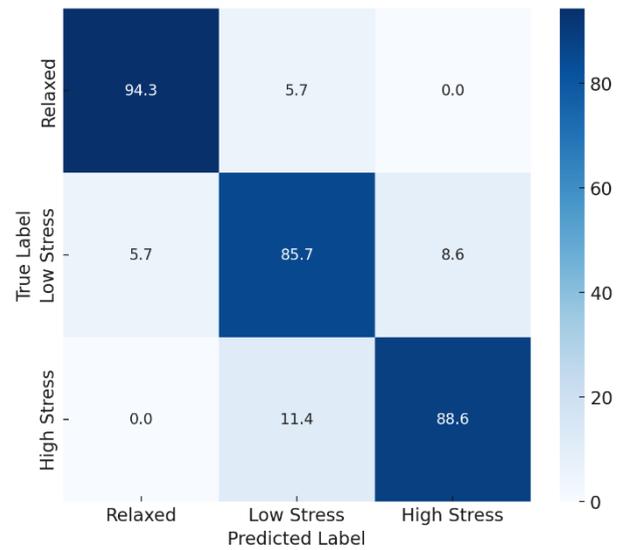

**Fig. 7** Confusion matrix of percentages for 3-class classification using alpha-TV-DTF with SVM.

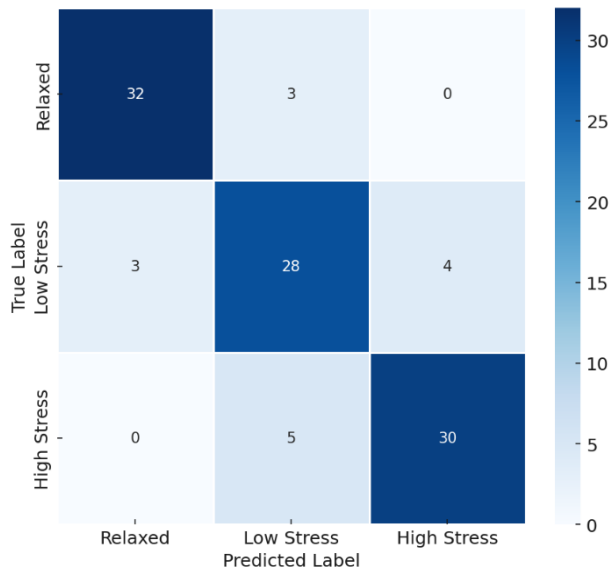

**Fig. 8** Confusion matrix of raw counts for 3-class classification using beta-TV-DTF with SVM.

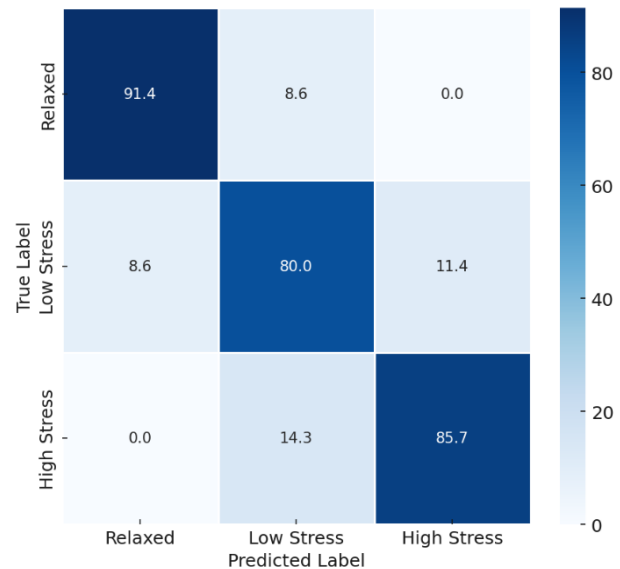

**Fig. 9** Confusion matrix of percentages for 3-class classification using beta-TV-DTF with SVM.

AdaBoost showed a similar rise, advancing from 54% F1 and 57% accuracy with beta-AP to 70% F1 and 72% accuracy with beta-TV-DTF. XGBoost demonstrated robust gains, rising from 56% F1 and 54% accuracy with beta-AP to 83% F1 and 86% accuracy with beta-TV-DTF, making it one of the strongest performers across all classifiers. These findings demonstrate that beta-band temporal connectivity measures are more discriminative than static power-based features, making TV-DTF a stronger choice for stress classification.

To complement the averaged cross-validation results reported in Table 2 to Table 4, confusion matrices are provided for alpha-TV-DTF and beta-TV-DTF using SVM in the 3-class classification scenario. Confusion matrices for alpha-TV-DTF with SVM are presented in terms of raw counts and normalized percentages.

The raw count matrix from Fig. 6 shows that the model correctly classified 94 out of 105 trials, yielding an accuracy of 89.5%, which closely aligns with the reported cross-validation mean of 89.73%. The percentage matrix from Fig. 7 further illustrates the per-class performance, with Relaxed identified at 94.3%, Low Stress at 85.7%, and High Stress at 88.6%. These results highlight that alpha-TV-DTF features lead to a more stable representation of stress classes, producing higher and more consistent outcomes across folds compared to other bands, and the agreement between raw counts and averaged percentages underscores the robustness of the SVM classifier with alpha-TV-DTF features.

Confusion matrices for beta-TV-DTF with SVM are also reported in terms of raw counts and percentages. The raw count matrix from Fig. 8 shows 90 correctly classified trials out of 105, yielding an accuracy of 85.7% accuracy, which is slightly lower than the cross-validation accuracy of 87.95%. The percentage matrix from Fig. 9 highlights class-specific performance, with Relaxed identified at 91.4%, Low Stress at 80.0%, and High Stress at 85.7%. Compared to alpha-TV-DTF, the variability across folds is higher for beta-TV-DTF, particularly with more confusion occurring between Low Stress and High Stress categories. This indicates that while beta-based effective connectivity features remain informative, they are more sensitive to fold-level differences and less stable than alpha-TV-DTF features for stress classification.

The classification outcomes described in Table 3 using alpha-TV-DTF and alpha-PLV shows that temporal connectivity features consistently outperform static functional connectivity features. From the results, SVM with alpha-TV-DTF achieved 85% precision, 90% recall, 88% F1 score, and 90% accuracy, while alpha-PLV with SVM produced 76% precision, 71% recall, 73% F1 score, and 75% accuracy. Similarly, RF models improved from 71% F1 and 73% accuracy with alpha-PLV to 74% F1 and 77% accuracy with alpha-TV-DTF. GB classifiers also reflected this gain, advancing from 68% F1 and 70% accuracy with alpha-PLV to 70% F1 and 71% accuracy with alpha-TV-DTF. AdaBoost reinforced the same trend, moving from 65% F1 and 66% accuracy with alpha-PLV to 59% F1 and 60% accuracy with alpha-TV-DTF, though with a narrower margin. XGBoost further strengthened the advantage of temporal modeling, improving from 76% F1 and 78% accuracy with alpha-PLV to 77% F1 and 79% accuracy with alpha-TV-DTF. These comparisons confirm that alpha-band temporal features capture stress-related dynamics more effectively.

Following the same trend, beta-TV-DTF features provided superior classification performance compared to beta-PLV across most machine learning models. For instance, SVM with beta-TV-DTF reached 88% precision, 80% recall, 84% F1 score, and 88% accuracy, while beta-PLV with SVM yielded 78% precision, 73% recall, 75% F1 score, and 77% accuracy. RF models improved from 74% F1 and 75% accuracy with beta-PLV to 75% F1 and 78% accuracy with beta-TV-DTF. GB classifiers also displayed a clear margin, with beta-TV-DTF achieving 69% F1 and 72% accuracy compared to beta-PLV's 71% F1 and 72% accuracy. AdaBoost continued the trend, moving from 67% F1 and 70% accuracy with beta-PLV to 70% F1 and 71% accuracy with beta-TV-DTF. XGBoost exhibited the strongest improvement, advancing from 78% F1 and 80% accuracy with beta-PLV to 83% F1 and 86% accuracy with beta-TV-DTF. These findings indicate that beta-band temporal connectivity features offer more discriminative insights into stress variation than static phase-locking measures.

The hyperparameters provided in Table 5 were used to generalize across both 2-class and 3-class stress classification scenarios using EEG-based TV-DTF features. SVM with an RBF kernel (C = 1, $\gamma$ = 0.05) to capture non-linear boundaries; RF with 200 trees and max_features = sqrt for stable variance reduction; GB (300 trees, learning rate = 0.05, depth = 6, subsample = 0.8) and AdaBoost (depth = 3 base tree, learning rate = 0.1) as complementary boosting baselines; and XGBoost (300 trees, learning rate = 0.05, depth = 6, subsample/colsample = 0.8, $\gamma$ = 0.1, $\lambda$ = 10) for regularized, high-capacity modeling. Because the dataset is imbalanced, we explicitly applied class reweighting for SVM such as class_weight = "balanced", RF class_weight = "balanced_subsample", and XGBoost used scale_pos_weight (ratio = neg/pos) for 2-class along with sample_weight; GB and AdaBoost were also trained with sample_weight derived from class weights each fold. Hyperparameter tuning and model evaluation were guided by macro-F1, which equally weights all classes and therefore enhances sensitivity to minority stress conditions, while accuracy was concurrently reported as a standard measure of overall performance.

Table 5 Hyperparameter setting used for all ML models

| Classifiers | Parameter/Hyperparameter | Values/Standards |
|---|---|---|
| SVM | Kernel | RBF |
|  | C (Regularization) | 1 |
|  | $\gamma$ (Kernel Coefficient) | 0.05 |
| RF | n_estimators | 200 |
|  | max_depth | 15 |
|  | min_samples_split | 5 |
|  | min_samples_leaf | 2 |
|  | max_features | sqrt |
| GB | n_estimators | 300 |
|  | learning_rate | 0.05 |
|  | max_depth | 6 |
|  | min_samples_split | 2 |
|  | min_samples_leaf | 2 |
|  | subsample | 0.8 |
| AdaBoost | n_estimators | 100 |
|  | learning_rate | 0.1 |
|  | base_estimator | DecisionTreeClassifier |
|  | max_depth | 3 |
| XGBoost | n_estimators | 300 |
|  | learning_rate | 0.05 |
|  | max_depth | 6 |
|  | subsample | 0.8 |
|  | colsample_bytree | 0.8 |
|  | $\gamma$ | 0.1 |
|  | $\lambda$ (L2 regularization) | 10 |

## Evaluation of Important TV-DTF Features

Given its superior performance and ability to handle complex interactions, XGBoost will also be employed in this research to evaluate the importance of TV-DTF features. XGBoost can be a powerful tool that can effectively identify the most critical features, offering significant insights into which temporal connectivity patterns contribute most to stress classification [43]. During the training process, XGBoost constructs an ensemble of DTs, each aimed at correcting errors from the previous trees [40] [43]. This method evaluates potential splits based on a metric known as gain, and features that result in higher gain are more frequently selected for splits [43]. By summing up the gains

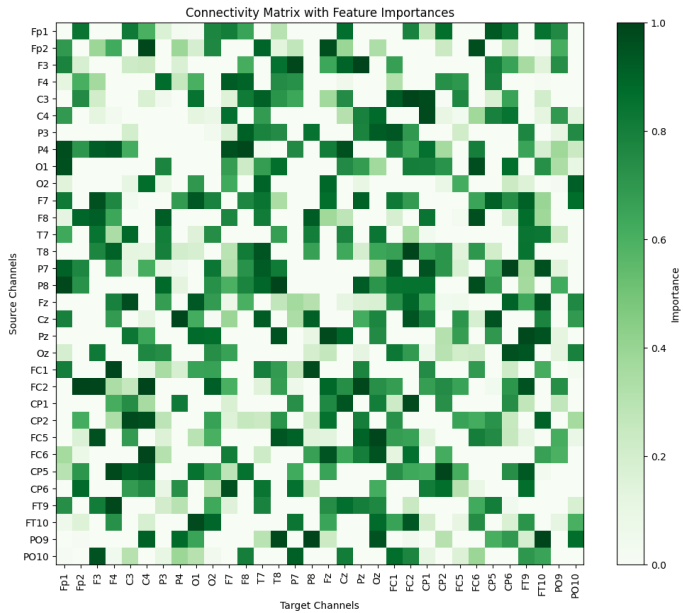

**Fig. 10** Feature matrix of alpha-TV-DTF with importance scoring.

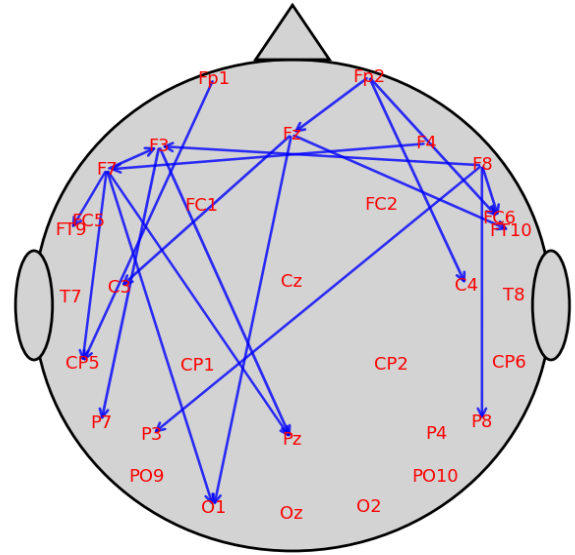

**Fig. 11** Directed influences of alpha-TV-DTF above importance score 0.9.

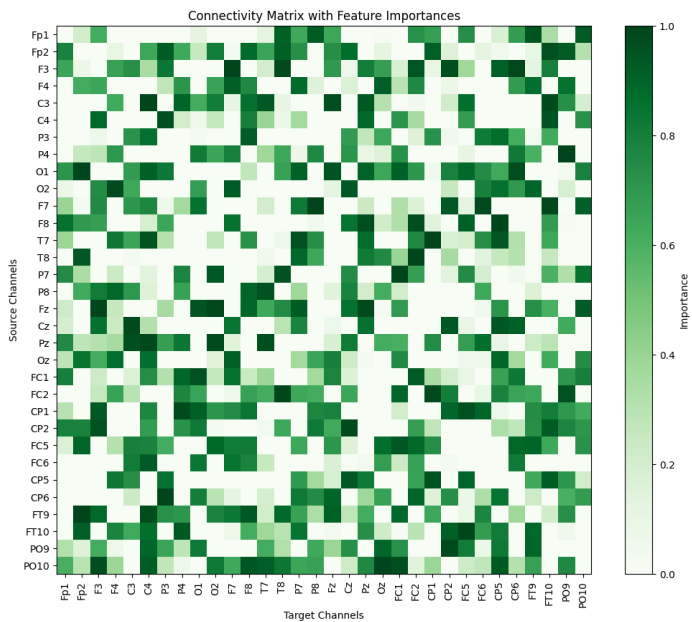

**Fig. 12** Feature matrix of beta-TV-DTF with importance scoring.

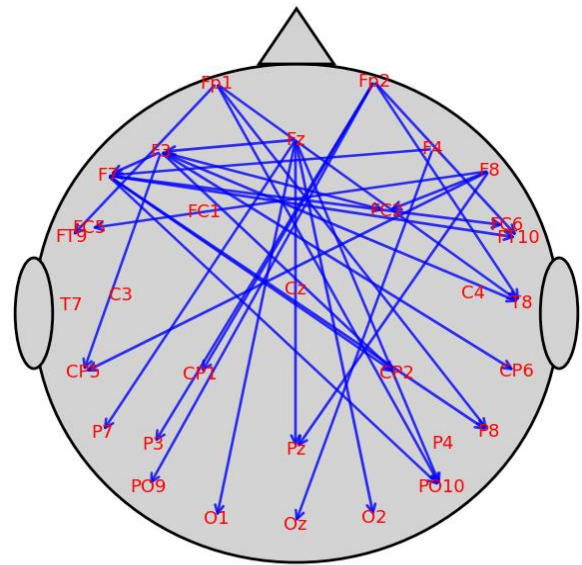

**Fig. 13** Directed influences of beta-TV-DTF above importance score 0.9.

associated with each feature, XGBoost can compute the importance score for every feature connection, normalizing the score between 0 and 1 [43]. This can be implemented by using XGBoost's built-in functionality, which offers a ranking of features according to their impact on the model's predictive performance [43]. The importance scoring technique not only identifies key feature connections relevant to stress classification but also enhances the transparency of the model's decision-making process, making it easier to understand how different connectivity features contribute to stress prediction [43]. An importance score threshold of 0.9 was applied in Fig. 11 and Fig. 13 to highlight the most critical alpha-TV-DTF and beta-TV-DTF features that contribute to stress-related brain network modulation, illustrating the dynamic evolution of directed influences in the alpha and beta bands

The feature connections of alpha-TV-DTF, along with their importance scores, are shown in Fig. 10. When confined to an importance score above 0.9, the topological representation in Fig. 11 illustrates dominant frontal–parietal and frontal–occipital pathways. The alpha-TV-DTF connectivity matrix indicates that frontal electrodes such as Fp1, Fp2, F3, F4, Fz, F8 emerge as dominant drivers, exerting a strong top-down influence on parietal sites such as Pz, P7, P8 and occipital sites such as O1, PO9, PO10, which primarily act as receivers. This directional pattern highlights the frontal cortex's regulatory role over posterior sensory and attentional systems, consistent with executive control and stress-related modulation. Notably, several long-range fronto-occipital influences such as F7 → O1, Fz → O1 stand out. Left frontal electrodes (Fp1, F3, FC5) primarily project to left parietal/occipital sites (P7, O1, Pz) and right frontal electrodes (F4, F8, FC6) connect more strongly with right posterior region (P8). This finding aligns with the literature, where frontal–occipital alpha connectivity is associated with attentional modulation and frontal–parietal alpha interactions are linked to cognitive control during stress processing.

The feature connections of beta-TV-DTF, along with their importance scores, are shown in Fig. 12. When confined to an importance score above 0.9, the topological representation in Fig. 13 highlights dense frontal–parietal and frontal–occipital interactions, with an even stronger emphasis on central integration compared to the alpha band. The beta-TV-DTF connectivity matrix reveals that frontal electrodes such as Fp1, Fp2, F3, F4, Fz, and F8 act as major drivers, projecting strong influences toward central and posterior sites, including CP1, CP2, Pz, O1, Oz, and PO10. These directional flows emphasize the frontal cortex's top-down regulation of parietal and occipital regions, consistent with the beta band's role in active cognitive control, motor readiness, and stress-related vigilance. Notably, strong long-range fronto-parietal-occipital influences such as Fp1 → PO10, Fp2 → PO9, and Fz → Pz are observed, suggesting executive modulation over visual-attentional systems. Additionally, Pz and PO10 emerge as major hubs, integrating bilateral frontal inputs and redistributing them across posterior regions. A hemispheric specialization is also evident, with left frontal nodes (Fp1, F3, FC5) influencing left posterior regions (O1, PO9), while right frontal sites (F4, F8, FC6) project to right-sided parietal-occipital areas (P8, PO10). This connectivity profile underscores the role of beta-band fronto-parietal-occipital pathways in maintaining heightened attentional control and sensorimotor readiness under stress.

Both alpha-TV-DTF and beta-TV-DTF connectivity features, as highlighted by XGBoost, demonstrate strong discriminative power in differentiating stress levels. Alpha-TV-DTF emphasizes dominant fronto-occipital influences, reflecting executive control over posterior attentional systems and making it a sensitive marker of stress-related connectivity modulation. In contrast, beta-TV-DTF reveals widespread frontal-driven connectivity toward parietal occipital and parietal hubs, consistent with heightened vigilance, cognitive control, and motor readiness under stress. The temporal resolution of TV-DTF further reveals how these directed interactions fluctuate across successive task windows, providing richer insight than static measures. The alignment of alpha modulation with top-down control and beta modulation with task-engagement dynamics highlights their neurophysiological relevance. Moreover, the convergence of these patterns across multiple classifiers confirms the stability and reliability of alpha and beta band temporal effective connectivity as biomarkers of stress.

## Impact of TV-DTF Features on ML models

Moving towards the impacts of the top 50 most important alpha-TV-DTF and beta-TV-DTF, the features were tested using the F1 score and accuracy metric by sequentially increasing the number of features from 5 to 50. This provides a comprehensive insight into how the performance of all ML models evolves as the number of features increases. The top 50 alpha-TV-DTF and beta-TV-DTF feature connections were selected for this analysis to strike a balance in the investigation of the performance in stress classification and feature complexity. The accuracy metric is used in this context since it indicates the overall correctness of the model, while the F1 score is used as it balances precision and recall.

Accordingly, from Fig. 14, the alpha-TV-DTF results suggest a consistent improvement for all models as the number of features increases, with SVM and XGBoost leading the performance in utilizing the top features to enhance their predictive performance in stress classification. Reaching F1 scores of around 70% demonstrates their effectiveness in capturing relevant stress-related connectivity patterns from alpha-TV-DTF features. GB shows stable improvement, achieving an F1 score close to 65%. In contrast, RF and AdaBoost exhibit the least improvement, with F1 scores peaking below 55% and 50%, respectively, indicating that these models are less adept at leveraging incremental features. Similarly, from an accuracy perspective, as shown in Fig. 15, the effect of the top 50 important alpha-TV-DTF results shows analogous trends, where both SVM and XGBoost again surpass the other models, achieving accuracies above 75%. RF and GB maintain moderate improvement, achieving around 70% accuracy, while AdaBoost reaches accuracy just above 60%.

Likewise, from Fig. 16, the beta-TV-DTF results suggest that the F1 scores of all models show a consistent improvement as more features are added. SVM and XGBoost lead the model performances, with SVM reaching an F1 score of above 70% and XGBoost closely following, demonstrating these models' efficiency in capturing stress-related connectivity patterns from the features. GB also shows improvement in achieving an F1 score above 65%. In contrast to other models, RF and AdaBoost exhibit the least improvement, with F1 scores peaking below 60% and above 60% respectively, indicating that these models are less adept at leveraging the incremental features. In terms of accuracy, as shown in Fig. 17, the model demonstrates identical trends with SVM and XGBoost, consistently outperforming the other models, achieving higher accuracy of around 80% and above 75% respectively. However, GB maintains a better improvement with accuracy peaking above 75%, while RF and AdaBoost achieve moderate improvements above 70%. For beta-TV-DTF, SVM and XGBoost show the most effective for enhanced stress classification.

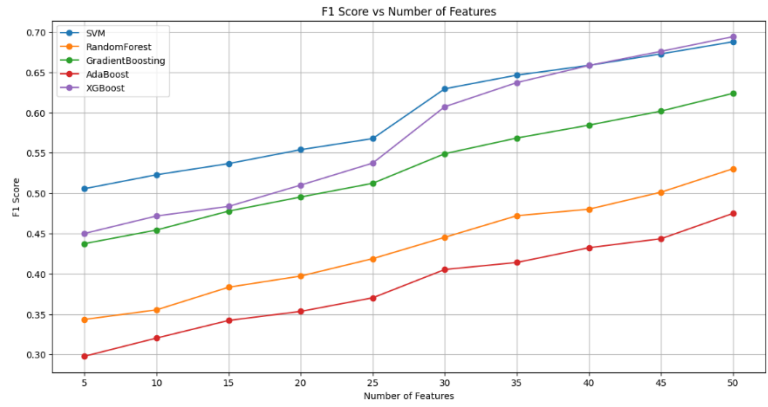

**Fig. 14** Impact of alpha-TV-DTF features on F1 scores in all ML models.

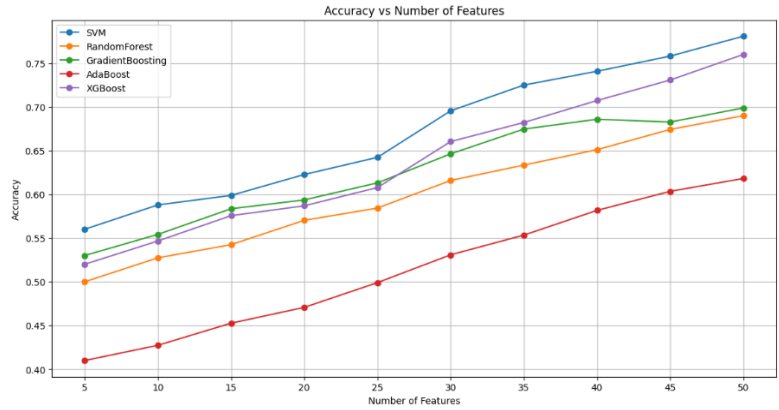

**Fig. 15** Impact of alpha-TV-DTF features on Accuracy in all ML models.

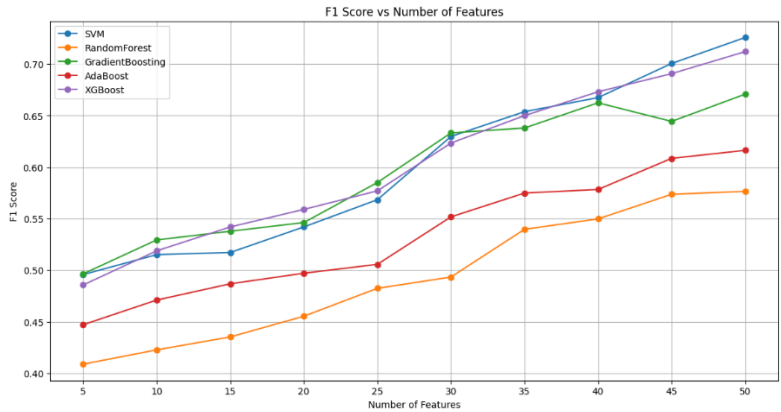

**Fig. 16** Impact of beta-TV-DTF features on F1 scores in all ML models.

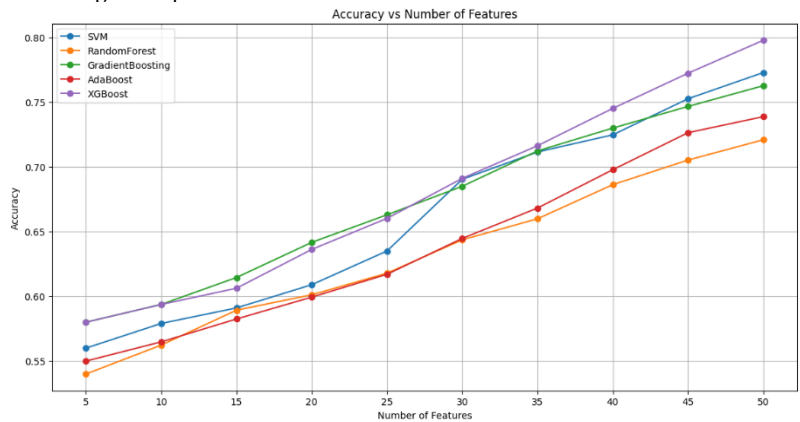

**Fig. 17** Impact of beta-TV-DTF features on Accuracy in all ML models.

## Discussion

In previous stress literature, variable findings exist regarding delta, theta, alpha, beta and gamma activity correlations with stress [44]. While some studies report increased delta activity, others noted a decrease during stress [45] [46] [47]. Similarly, theta, alpha and beta activity have shown both increases and decreases in stressed situations [13] [14] [23] [45] [46] [47] [48] [49]. Although one study found increased gamma activity during stress [48], caution is warranted due to potential myogenic activity interferences [44] [48].

To comprehend the discriminative power of temporal connectivity features, XGBoost was employed to evaluate the importance of TV-DTF connections, while enabling the identification of the most critical pathways contributing to stress classification. The feature importance analysis from Fig. 10 revealed that alpha-TV-DTF was dominated by frontal-driven connections projecting toward parietal and occipital regions, with strong long-range fronto-occipital influences highlighting the executive control of attentional systems in Fig. 11. In contrast, beta-TV-DTF exhibited a denser frontal–parietal–occipital interaction with central hubs such as Pz and PO10 in Fig. 13, reflecting heightened vigilance, motor readiness, and cognitive control under stress. Together, these results demonstrate that alpha and beta bands capture complementary aspects of stress-related network dynamics, where alpha reflects regulatory modulation of posterior regions, while beta emphasizes integrative control and sensorimotor preparedness. This ML-based feature understanding not only strengthens the interpretability of the classification framework but also reinforces TV-DTF's role as a robust marker of stress-induced brain connectivity.

The impact of the top 50 most important alpha-TV-DTF and beta-TV-DTF features on ML models from Fig. 14 to Fig. 17 further validates its discriminative capacity toward stress classification. As the number of features increased from 5 to 50, all models demonstrated gradual improvements in F1 score and accuracy, with SVM, GB and XGBoost consistently outperforming the others. For alpha-TV-DTF, these models achieved F1 scores near 70% and accuracies above 75%, while GB showed moderate gains, RF and AdaBoost lagged with limited improvement. A similar trend was observed for beta-TV-DTF, where SVM surpassed 70% F1 and reached near 80% accuracy, and XGBoost followed closely, demonstrating their efficiency in leveraging temporal connectivity features. Although GB achieved a competitive accuracy above 75%, RF and AdaBoost again showed weaker improvements. These results highlight that while incremental features improve classification across all models, especially SVM and XGBoost, are most effective at exploiting the top alpha-TV-DTF and beta-TV-DTF connections, strengthening their suitability for modelling stress-related brain connectivity.

The classification outcomes further validate the effectiveness of TV-DTF features as robust markers for stress-related brain dynamics. While delta-TV-DTF and theta-TV-DTF features showed only modest predictive power, particularly with SVM as the strongest performer, their discriminability was limited compared to higher-frequency bands. In contrast, alpha-TV-DTF and beta-TV-DTF features consistently delivered superior classification results across multiple models, with SVM and XGBoost achieving the highest accuracies. In particular, alpha connectivity patterns reached near 90% accuracy, reflecting the importance of frontal-driven alpha pathways in attentional control under stress. Beta connectivity also provided strong discriminability, capturing the heightened vigilance and cognitive control processes associated with stress. These findings highlight that temporal effective connectivity in the alpha and beta bands provides the most reliable features for stress classification. However, the classification results were required to compare with traditional PSD-based outcomes to further validate the dynamic temporal effective connectivity features and demonstrate that TV-DTF captures dynamic neural interactions beyond static spectral power.

Across all models, alpha-TV-DTF and beta-TV-DTF significantly outperformed their AP counterparts, with improvements most evident in SVM, RF, AdaBoost and XGBoost, where directional temporal features enhanced classification accuracy by 20–30%. This demonstrates that static spectral power is insufficient to fully capture stress-induced neural dynamics, whereas TV-DTF effectively models the temporal flow of information across brain regions. Overall, these results provide an effective ML-based validation where lightweight TV-DTF features offer a more discriminative and generalizable framework for EEG-based stress classification. Such findings open the way for real-time stress monitoring systems that leverage temporal connectivity features for reliable detection in practical applications.

To enable a fair comparison with previous EEG-based stress studies, which mostly report binary stress classification (Relaxed vs Stress), both 2-class and 3-class scenarios were evaluated from Fig. 18 to Fig. 21. This comparison demonstrates that our framework not only generalizes across traditional binary setups, but also capture the finer granularity of multi-level stress classification. Interestingly, the classification performance of alpha-band connectivity features was higher in the 3-class scenario compared to the 2-class setup. Although this may appear counterintuitive, it reflects the distinctive role of alpha oscillations, which are strongly associated with attentional control, relaxation, and cognitive effort.

When Low and High Stress conditions collapse into a single binary stress class, the discriminative alpha patterns between them are lost, reducing separability. In contrast, the 3-class setup preserves these graded differences, enabling classifiers such as SVM with nonlinear kernels to exploit the structured boundaries, where Low Stress effectively serves as a buffer zone between Relaxed and High Stress. Moreover, the 2-class split is more imbalanced than the 3-class division, which further depresses binary performance. These findings suggest that Alpha connectivity is particularly sensitive to fine-grained stress distinctions and that the inclusion of intermediate classes can improve generalization by preserving the intrinsic variability of brain dynamics under stress. In contrast, the other frequency bands followed the more typical pattern in which binary classification yielded higher performance than multiclass, underscoring the unique sensitivity of Alpha-band connectivity to graded stress variations.

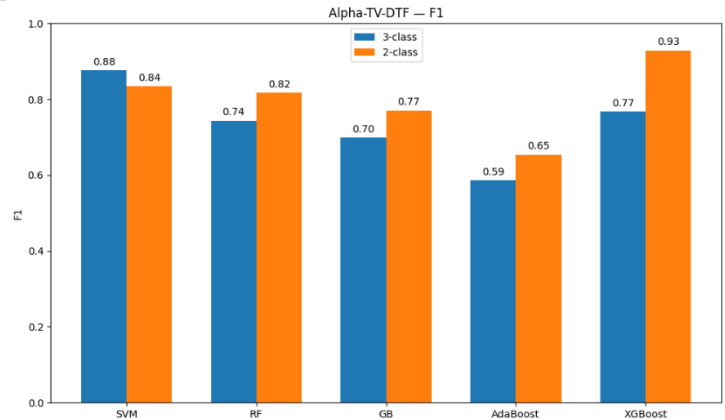

**Fig. 18** Comparison of alpha-TV-DTF-based 2-class and 3-class classification on F1 scores in all ML models.

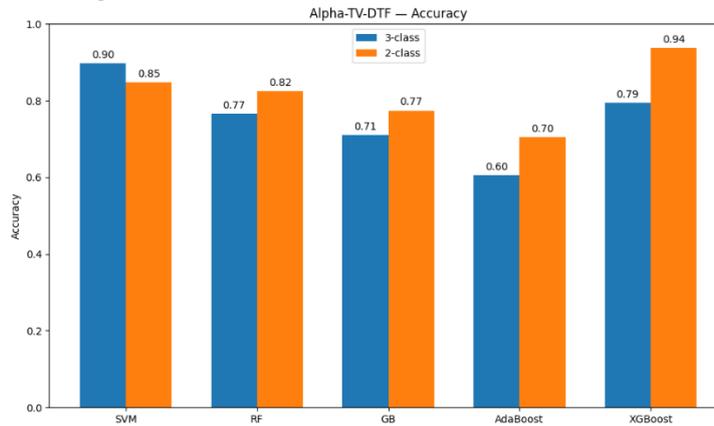

**Fig. 19** Comparison of alpha-TV-DTF-based 2-class and 3-class classification on Accuracy in all ML models.

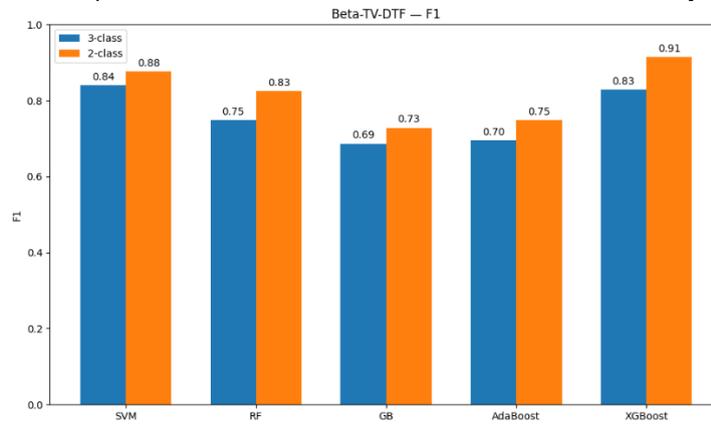

**Fig. 20** Comparison of beta-TV-DTF-based 2-class and 3-class classification on F1 scores in all ML models.

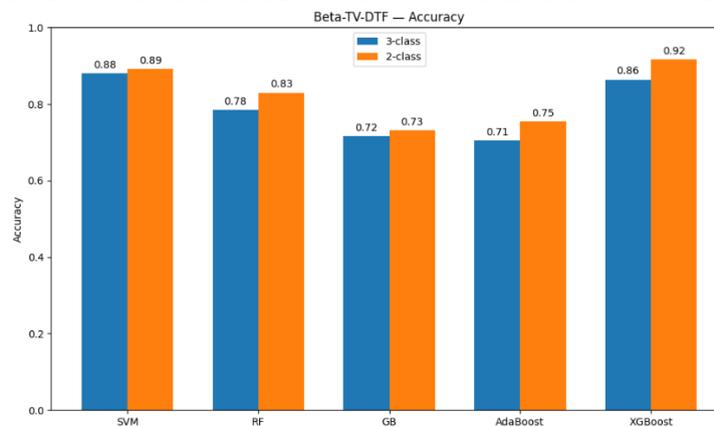

**Fig. 21** Comparison of beta-TV-DTF-based 2-class and 3-class classification on Accuracy in all ML models.

Table 6 Comparison table of stress studies based on connectivity features.

| Ref (year) | Experimental Factors | | | | Connectivity Features | Key electrodes | Top Classifier (Class) | Best Accuracy Achieved (Feature, Class) | Observations |
|---|---|---|---|---|---|---|---|---|---|
| | Dataset used | Stress tasks | No. of EEG Channels | No. of Subjects | | | | | |
| [13] Alonso et al. (2015) | Custom | MAT | 19 | 30 | Coherence, MSCE, CMIF. | Not Identified | Not Classified | Not Classified | Reports scalp connectivity maps using coherence and CMIF with significant regional patterns, high-beta coherence increases; alpha coherence decreases in anterior regions. |
| [18] Khosrowabadi et al. (2018) | Custom | Long-term examination | 8 | 26 | DTF, gPDC, PSI. | C3, C4, F4, P3, P4, T8 | SVM (4) | 90.90% (PSI, 4) | PSI revealed hubs such as C4 dominant in positive emotion; P3 dominant in negative emotion |
| [21] Balconi et al. (2018) | Custom | Competition/ social stress | 15 | 14 | Correlation-based inter-brain connectivity. | Not Identified | Not Classified | Not Classified | Partial correlation leading to undirected functional connectivity. |
| [12] Darzi et al. (2019) | Custom | IAPS + audio clips | 8 | 26 | MSCE, GC, PSI, DTF. | P3 and T8 | SVM (2) | Above 90% (GC, 2) Above 90% (DTF, 2) | The P3 to T8 connection was reported as the strongest and most discriminative feature. |
| [25] Hag et al. (2021) | Custom | MAT | 7 | 22 | Delta-PLV, Theta-PLV, Alpha-PLV, Sigma-PLV, Low beta-PLV, High beta-PLV. | F7, F3, Fz, F8, Fp1, F4 | LDA (2) | 75.20% (Delta-PLV, 2), 71.90% (Alpha-PLV, 2), 73.40 % (High beta-PLV, 2). | PLV is an undirected measure; although the study reported significant frontal channels, it did not identify directional information flow. |
| [50] Al-Shargie et al. (2021) | Custom | MAT | 7 | 25 | Alpha-based Coherence | F3, F4, F7, F8, Fp1 and Fp2. | Not Classified | Not Classified | Reports the dominance of right dorsolateral prefrontal cortex in alpha band. |
| [26] Vanhollebeke et al. (2023) | Custom | Visual reasoning puzzles | 57 | 73 | Alpha-based power, Amplitude Envelope Correlation (AEC) | Not Identified | Not Classified | Not Classified | Reports Significant alpha power increases in left precuneus, right precuneus, right posterior cingulate cortex |
| Proposed Study | SAM 40 | MAT | 32 | 35 | Delta-TV-DTF, Theta-TV-DTF, Alpha-TV-DTF, Beta-TV-DTF, Gamma-TV-DTF, | Alpha: frontal (F3, F4, Fz, F7, F8); parietal/occipital (P3, P4, Pz, O1, O2). Beta: central/parietal (Cz, Pz, PO10) | XGBoost (2), SVM (3) | 93.69% (Alpha-TV-DTF, 2); 89.73% (Alpha-TV-DTF, 3) | Reports alpha connectivity with long-range frontal→parietal and frontal→occipital connections. Beta connectivity with denser frontal–parietal–occipital interactions with hubs around central and parietal sites. |

The 'Dataset Used' column in Table 6 differentiates between publicly accessible datasets, such as the SAM 40 dataset [27], and custom datasets collected specifically for the respective studies. Although these custom datasets are not publicly available, they were generated using well-established, replicable task-based protocols such as arithmetic tasks, Stroop tasks, or audio–visual stimuli. Consequently, similar datasets can be reproduced under comparable experimental settings, which helps to partially offset the limitations arising from restricted data accessibility. In the EEG-based stress classification studies listed in Table 6, it can be observed that previously connectivity features such as GC and DTF have achieved above 90% accuracy in classifying stress in a 2-class while using 8 EEG channels among 26 participants' data [12]. In contrast, the PSI feature has achieved 90.90% accuracy in classifying stress on a 4-class while using 8 EEG channels among 26 participants' data [18]. However, these connectivity features lack directional influence with the frequency specificity, indicating that the features do not provide any insight into the stress-affected informational influence among brain regions at the EEG frequency bands. This limitation could lead to overlooking crucial data on brain frequency-specific information flow during periods of stress, which is essential to EEG studies. Furthermore, one study used 7 EEG channels among 22 participants and observed 75.20%, 71.90% and 73.40% accuracy in 2-class stress classification while using frequency-specific functional connectivity features, namely PLV in delta, alpha and high beta, respectively [25]. Nevertheless, PLV provides the phase synchronization of EEG channels rather than directional influences. Similar to PLV, researchers have implemented AEC, a functional connectivity measure [26], which provides a statistical dependency among the brain regions but does not provide any information about frequency-specificity and time-varying directional influences of the brain dynamics. Moreover, the TV-DTF features of the proposed framework not only provide meaningful understandings of frequency-specificity but also provide the time-varying directional influences of information flow while achieving higher classification accuracy of 93.69% while implementing XGBoost in 2-class, and 89.73% while implementing SVM in 3-class mental stress classification, both by utilizing alpha-TV-DTF features.

### Advantages and limitations

The proposed lightweight TV-DTF framework demonstrates several advantages discussed in this subsection. The framework captures the temporal, directional and informational stress-related brain patterns by modelling dynamic effective connectivity between EEG electrodes, which the static metrics such as PSD, PLV, and coherence could not provide. Second, the proposed framework is designed to be lightweight and scalable, with a reproducible MATLAB-based implementation that reduces runtime complexity compared to traditional DTF, as described in Fig. 4. Nextly, the method generalizes across both 2-class and 3-class stress scenarios, achieving 89.73% accuracy in 3-class and 93.69% accuracy in 2-class using alpha-TV-DTF features with SVM and XGboost, respectively, thereby ensuring robustness. At the same time, it offers strong neurological interpretability, as the feature importance analysis in Fig. 11 and Fig. 13 revealed dominant frontal–parietal and frontal–occipital pathways, underscoring the regulatory role of the frontal cortex in stress processing and providing physiological meaning beyond classification. Lastly, using subject-wise 10-fold cross-validation prevents overlap between training and test data, thereby avoiding data leakage and strengthening the reliability of the lightweight TV-DTF frameworks' results.

On the other hand, despite these strengths, there are certain limitations in the proposed methodology. The research study was restricted to the SAM 40 dataset with 35 young, healthy participants, which may limit generalizability across clinical populations. Additionally, while alpha and beta bands yielded strong discriminative power, delta, theta, and gamma bands showed weaker performance, which suggests EEG band-specific variability. This framework also depends on fixed EEG segmentation parameters, such as a 5-second window and 2-second overlap, which may not optimally capture the brain dynamics across other stress tasks. Finally, this validation was conducted in a controlled offline setting using the SAM 40 dataset; therefore, further research is needed to extend the lightweight framework to real-time stress monitoring scenarios involving more diverse stress-inducing paradigms.

## Conclusion

In this study, the SAM 40 dataset was employed to validate a lightweight framework proposed for Time-Varying Directed Transfer Function (TV-DTF) computation, where dynamic TV-DTF features across multiple EEG frequency bands were utilized for ML-based mental stress classification. The ML-based validation revealed that delta-TV-DTF and theta-TV-DTF features provided modest discriminability, while alpha-TV-DTF and beta-TV-DTF features consistently demonstrated better performances with SVM and XGBoost, emerging as the most reliable classifiers. In particular, alpha-TV-DTF achieved a near 93.69% accuracy, while beta-TV-DTF delivered high accuracies of 91.73% both with XGBoost in 2-class, confirming the strong discriminative power of higher-frequency connectivity. When TV-DTF features were compared to PSD and PLV-based features, alpha-TV-DTF and beta-TV-DTF consistently outperformed the other measures across all models, establishing the robustness of the temporal effective connectivity framework in stress detection. Furthermore, feature importance analysis with XGBoost identified dominant long-range fronto-occipital influences for alpha-TV-DTF and long-range fronto-parietal-occipital influences for beta-TV-DTF. It shows that frontal electrodes acted as primary drivers exerting top-down regulation on posterior regions, reflecting executive control, attentional modulation, and vigilance under stress. Finally, the impact of the top important features demonstrated that incremental connectivity features enhanced the model performances, with SVM and XGBoost effectively leveraging alpha-TV-DTF and beta-TV-DTF connections to achieve robust classification. These findings validate the lightweight TV-DTF framework as a powerful tool for EEG-based stress quantification, offering improved accuracy and interpretability across diverse stress levels.

**Acknowledgements** I would like to express my sincere appreciation to Samer Hanoun and Imali Hettiarachchi from IISRI of Deakin University for their valuable support during the planning and early development of this research.

**Author Contribution** All authors contributed to the study conception and design. Material preparation, data analysis, and visualization were carried out collaboratively by Sayantan Acharya, Abbas Khosravi, Douglas Creighton, Roohallah Alizadehsani, and U. Rajendra Acharya. The first draft of the manuscript was prepared by Sayantan Acharya, and all authors provided feedback on earlier versions. All authors read and approved the final manuscript.

**Data Availability** The data that support the findings of this study are openly available in the SAM 40 database at https://www.sciencedirect.com/science/article/pii/S2352340921010465, referenced as [27].

## Declarations

**Ethical Approval** Not applicable.

**Informed Consent** Not applicable.

**Conflict of Interest** The authors declare no competing interests.

## References

1. Arsalan A, Majid M, Nizami IF, Manzoor W, Anwar SM, Ryu J. Human stress assessment: A comprehensive review of methods using wearable sensors and non-wearable techniques. arXiv preprint arXiv:2202.03033. 2022 Feb 7.
2. Kivimäki M, Steptoe A. Effects of stress on the development and progression of cardiovascular disease. Nature Reviews Cardiology. 2018 Apr;15(4):215-29.
3. Shalev A, Liberzon I, Marmar C. Post-traumatic stress disorder. New England journal of medicine. 2017 Jun 22;376(25):2459-69.
4. Sinha R. Stress and substance use disorders: risk, relapse, and treatment outcomes. The Journal of clinical investigation. 2024 Aug 15;134(16).
5. Chan SF, La Greca AM. Perceived stress scale (PSS). InEncyclopedia of behavioral medicine 2020 Oct 20 (pp. 1646-1648). Cham: Springer International Publishing.
6. Al-Shargie F, Tang TB, Kiguchi M. Stress assessment based on decision fusion of EEG and fNIRS signals. IEEE Access. 2017 Sep 19;5:19889-96.
7. Ahn JW, Ku Y, Kim HC. A novel wearable EEG and ECG recording system for stress assessment. Sensors. 2019 Apr 28;19(9):1991.
8. Affanni A. Wireless sensors system for stress detection by means of ECG and EDA acquisition. Sensors. 2020 Apr 4;20(7):2026.
9. Ali N, Nater UM. Salivary alpha-amylase as a biomarker of stress in behavioral medicine. International journal of behavioral medicine. 2020 Jun;27(3):337-42.
10. Fischer T, Halmerbauer G, Meyr E, Riedl R. Blood pressure measurement: a classic of stress measurement and its role in technostress research. InInformation Systems and Neuroscience: Gmunden Retreat on NeuroIS 2017 2017 Nov 17 (pp. 25-35). Cham: Springer International Publishing.
11. McEwen BS, Akil H. Revisiting the stress concept: implications for affective disorders. Journal of Neuroscience. 2020 Jan 2;40(1):12-21.
12. Darzi A, Azami H, Khosrowabadi R. Brain functional connectivity changes in long-term mental stress. Journal of Neurodevelopmental Cognition. 2022 Apr 9;1(1):16-41.
13. Alonso JF, Romero S, Ballester MR, Antonijoan RM, Mañanas MA. Stress assessment based on EEG univariate features and functional connectivity measures. Physiological measurement. 2015 May 27;36(7):1351.
14. Al-Shargie FM, Tang TB, Badruddin N, Kiguchi M. Mental stress quantification using EEG signals. InInternational conference for innovation in biomedical engineering and life sciences 2015 Dec 6 (pp. 15-19). Singapore: Springer Singapore.
15. Wang XW, Nie D, Lu BL. Emotional state classification from EEG data using machine learning approach. Neurocomputing. 2014 Apr 10;129:94-106.
16. Ameera A, Saidatul A, Ibrahim Z. Analysis of EEG spectrum bands using power spectral density for pleasure and displeasure state. InIOP conference series: Materials science and engineering 2019 Jun 1 (Vol. 557, No. 1, p. 012030). IOP Publishing.
17. Attallah O. An effective mental stress state detection and evaluation system using minimum number of frontal brain electrodes. Diagnostics. 2020 May 9;10(5):292.
18. Khosrowabadi R. Stress and perception of emotional stimuli: Long-term stress rewiring the brain. Basic and clinical neuroscience. 2018 Mar;9(2):107.
19. van den Heuvel MP, Sporns O. A cross-disorder connectome landscape of brain dysconnectivity. Nature reviews neuroscience. 2019 Jul;20(7):435-46.
20. Subhani AR, Malik AS, Kamil N, Saad MN. Difference in brain dynamics during arithmetic task performed in stress and control conditions. In2016 IEEE EMBS Conference on Biomedical Engineering and Sciences (IECBES) 2016 Dec 4 (pp. 695-698). IEEE.

21. Balconi M, Vanutelli ME. Functional EEG connectivity during competition. BMC neuroscience. 2018 Oct 18;19(1):63.
22. Rho G, Callara AL, Cecchetto C, Vanello N, Scilingo EP, Greco A. Valence, arousal, and gender effect on olfactory cortical network connectivity: A study using dynamic causal modeling for EEG. IEEE Access. 2022 Dec 1;10:127313-27.
23. Al-Shargie F, Tang TB, Kiguchi M. Assessment of mental stress effects on prefrontal cortical activities using canonical correlation analysis: an fNIRS-EEG study. Biomedical optics express. 2017 Apr 19;8(5):2583-98.
24. Bastos AM, Schoffelen JM. A tutorial review of functional connectivity analysis methods and their interpretational pitfalls. Frontiers in systems neuroscience. 2016 Jan 8;9:175.
25. Hag A, Handayani D, Pillai T, Mantoro T, Kit MH, Al-Shargie F. EEG mental stress assessment using hybrid multi-domain feature sets of functional connectivity network and time-frequency features. Sensors. 2021 Sep 20;21(18):6300.
26. Vanhollebeke G, Kappen M, De Raedt R, Baeken C, van Mierlo P, Vanderhasselt MA. Effects of acute psychosocial stress on source level EEG power and functional connectivity measures. Scientific reports. 2023 May 31;13(1):8807.
27. Ghosh R, Deb N, Sengupta K, Phukan A, Choudhury N, Kashyap S, Phadikar S, Saha R, Das P, Sinha N, Dutta P. SAM 40: Dataset of 40 subject EEG recordings to monitor the induced-stress while performing Stroop color-word test, arithmetic task, and mirror image recognition task. Data in Brief. 2022 Feb 1;40:107772.
28. Allen AP, Kennedy PJ, Dockray S, Cryan JF, Dinan TG, Clarke G. The trier social stress test: principles and practice. Neurobiology of stress. 2017 Feb 1;6:113-26.
29. Wang W, Sun W. Using EEG effective connectivity based on granger causality and directed transfer function for emotion recognition. International Journal of Advanced Computer Science and Applications. 2023;14(9).
30. Acharya S, Khosravi A, Creighton D. Neural Dynamics Under Stress: Advanced Statistical Validation and Correlation of EEG-Based Time-Varying Effective Connectivity Features. In2025 25th International Conference on Digital Signal Processing (DSP) 2025 Jun 25 (pp. 1-5). IEEE.
31. Kostoglou K, Müller-Putz GR, Graz B. Optimizing time-varying autoregressive models for BCI applications. In9th Graz Brain-Computer Interface Conference: Join Forces-Increase Performance: BCI 2024 2024.
32. Bagherzadeh S, Maghooli K, Shalbaf A, Maghsoudi A. Recognition of emotional states using frequency effective connectivity maps through transfer learning approach from electroencephalogram signals. Biomedical Signal Processing and Control. 2022 May 1;75:103544.
33. Aho K, Derryberry D, Peterson T. Model selection for ecologists: the worldviews of AIC and BIC. Ecology. 2014 Mar 1;95(3):631-6.
34. Privalsky V. Multivariate Time and Frequency Domain Analysis. InTime Series Analysis in Climatology and Related Sciences 2020 Nov 23 (pp. 221-237). Cham: Springer International Publishing.
35. Cui, J., Xu, L., Bressler, S.L., Ding, M. and Liang, H., 2008. BSMART: a Matlab/C toolbox for analysis of multichannel neural time series. *Neural Networks*, 21(8), pp.1094-1104.
36. Mullen T. Source information flow toolbox (SIFT). Swartz Center Comput Neurosci. 2010 Dec 15;15:1-69.
37. Khan DM, Yahya N, Kamel N, Faye I. A novel method for efficient estimation of brain effective connectivity in EEG. Computer Methods and Programs in Biomedicine. 2023 Jan 1;228:107242.
38. Zhang Y, Ji X, Zhang S. An approach to EEG-based emotion recognition using combined feature extraction method. Neuroscience letters. 2016 Oct 28;633:152-7.
39. Nasteski V. An overview of the supervised machine learning methods. Horizons. b. 2017 Dec 15;4(51-62):56.
40. Rahman AA, Siraji MI, Khalid LI, Faisal F, Nishat MM, Islam MR. Detection of mental state from EEG signal data: an investigation with machine learning classifiers. In2022 14th International Conference on Knowledge and Smart Technology (KST) 2022 Jan 26 (pp. 152-156). IEEE.
41. Akella A, Singh AK, Leong D, Lal S, Newton P, Clifton-Bligh R, Mclachlan CS, Gustin SM, Maharaj S, Lees T, Cao Z. Classifying multi-level stress responses from brain cortical EEG in nurses and non-health professionals using machine learning auto encoder. IEEE Journal of Translational Engineering in Health and Medicine. 2021 May 5;9:1-9.
42. Awais MA, Yusoff MZ, Khan DM, Yahya N, Kamel N, Ebrahim M. Effective connectivity for decoding electroencephalographic motor imagery using a probabilistic neural network. Sensors. 2021 Sep 30;21(19):6570.
43. Hsieh CP, Chen YT, Beh WK, Wu AY. Feature selection framework for XGBoost based on electrodermal activity in stress detection. In2019 IEEE International Workshop on Signal Processing Systems (SiPS) 2019 Oct 20 (pp. 330-335). IEEE.
44. Giannakakis G, Grigoriadis D, Giannakaki K, Simantiraki O, Roniotis A, Tsiknakis M. Review on psychological stress detection using biosignals. IEEE transactions on affective computing. 2019 Jul 9;13(1):440-60.
45. Giannakakis G, Grigoriadis D, Tsiknakis M. Detection of stress/anxiety state from EEG features during video watching. In2015 37th Annual International Conference of the IEEE Engineering in Medicine and Biology Society (EMBC) 2015 Aug 25 (pp. 6034-6037). IEEE.
46. Bosl WJ, Bosquet Enlow M, Lock EF, Nelson CA. A biomarker discovery framework for childhood anxiety. Frontiers in Psychiatry. 2023 Jul 17;14:1158569.
47. Chang H, Zong Y, Zheng W, Xiao Y, Wang X, Zhu J, Shi M, Lu C, Yang H. EEG-based major depressive disorder recognition by selecting discriminative features via stochastic search. Journal of Neural Engineering. 2023 Mar 23;20(2):026021.
48. Minguillon J, Lopez-Gordo MA, Pelayo F. Stress assessment by prefrontal relative gamma. Frontiers in computational neuroscience. 2016 Sep 22;10:101.
49. Acharya S, Khosravi A, Creighton D, Alizadehsani R, Acharya UR. Neurostressology: A systematic review of EEG-based automated mental stress perspectives. Information Fusion. 2025 May 27:103368.
50. Al-Shargie F. Prefrontal cortex functional connectivity based on simultaneous record of electrical and hemodynamic responses associated with mental stress. arXiv preprint arXiv:2103.04636. 2021 Mar 8.